\documentclass[sigconf]{acmart}

\AtBeginDocument{%
  }

\setcopyright{acmlicensed}
\copyrightyear{2018}
\acmYear{2018}
\acmDOI{XXXXXXX.XXXXXXX}

\acmConference[Conference acronym 'XX]{Make sure to enter the correct
  conference title from your rights confirmation emai}{June 03--05,
  2018}{Woodstock, NY}
\acmISBN{978-1-4503-XXXX-X/18/06}
\usepackage{graphicx}
\usepackage{wrapfig}
\usepackage{listings}
\usepackage{xcolor}
\usepackage{multirow}
\usepackage{booktabs}

\newcommand{\revision}{}

\copyrightyear{2025}
\acmYear{2025}
\acmConference[CHI '25]{CHI Conference on Human Factors in Computing Systems}{April 26-May 1, 2025}{Yokohama, Japan}
\acmBooktitle{CHI Conference on Human Factors in Computing Systems (CHI '25), April 26-May 1, 2025, Yokohama, Japan}\acmDOI{10.1145/3706598.3713772}
\acmISBN{979-8-4007-1394-1/25/04}

\begin{document}



\title{Advancing Problem-Based Learning with Clinical Reasoning for Improved Differential Diagnosis in Medical Education}
\author{Yuansong Xu}
\orcid{0009-0005-1630-6279}
\affiliation{%
  \institution{ShanghaiTech University}
  \city{Shanghai}
  \country{China}
}
\email{xuys2023@shanghaitech.edu.cn}

\author{Yuheng Shao}
\orcid{0009-0008-6991-6427}
\affiliation{%
  \institution{ShanghaiTech University}
  \city{Shanghai}
  \country{China}
}
\email{shaoyh2024@shanghaitech.edu.cn}

\author{Jiahe Dong}
\orcid{0009-0003-1537-0717}
\affiliation{%
  \institution{ShanghaiTech University}
  \city{Shanghai}
  \country{China}
}
\email{dongjh@shanghaitech.edu.cn}

\author{Shaohan Shi}
\orcid{0009-0004-3384-8304}
\affiliation{%
  \institution{ShanghaiTech University}
  \city{Shanghai}
  \country{China}
}
\email{shishh2023@shanghaitech.edu.cn}

\author{Chang Jiang}
\affiliation{
  \institution{Shanghai Clinical Research and Trial Center}
  \city{Shanghai}
  \country{China}
}
\email{cjiang_fdu@yeah.net}
\authornotemark[1]

\author{Quan Li}
\affiliation{
  \institution{ShanghaiTech University}
  \city{Shanghai}
  \country{China}
}
\email{liquan@shanghaitech.edu.cn}
\authornotemark[0]
\authornote{Quan Li and Chang Jiang are the corresponding authors.}

\renewcommand{\shortauthors}{Xu et al.}

\begin{abstract}
Medical education increasingly emphasizes students' ability to apply knowledge in real-world clinical settings, focusing on evidence-based clinical reasoning and differential diagnoses. Problem-based learning (PBL) addresses traditional teaching limitations by embedding learning into meaningful contexts and promoting active participation. However, current PBL practices are often confined to medical instructional settings, limiting students' ability to self-direct and refine their approaches based on targeted improvements. Additionally, the unstructured nature of information organization during analysis poses challenges for record-keeping and subsequent review. Existing research enhances PBL realism and immersion but overlooks the construction of logic chains and evidence-based reasoning. To address these gaps, we designed \textit{e-MedLearn}, a learner-centered PBL system that supports more efficient application and practice of evidence-based clinical reasoning. Through controlled study (N=\revision{19}) and testing interviews (N=\revision{13}), we gathered data to assess the system's impact. The findings demonstrate that \textit{e-MedLearn} improves PBL experiences and provides valuable insights for advancing clinical reasoning-based learning.
\end{abstract}


\maketitle

\section{Introduction}
\par Modern medical education emphasizes students' ability to utilize their knowledge to reason and solve problems in real-world contexts, as well as fostering self-directed and continuous learning skills~\cite{kiesewetter2016knowledge,buja2019medical,knowles1975self,tagawa2008physician, Bordage2009ContinuingME}. However, traditional teaching methods tend to be ``\textit{teacher-centered and lecture-focused}''~\cite{ozturk2008comparison, murphy2021teacher}, with students passively receiving information and focusing on memorization and repetition. This approach may hinder the development of students' active learning and practical problem-solving skills. To facilitate students' active learning and usage of knowledge in real clinical scenarios, Problem-Based Learning (PBL) was proposed and developed~\cite{barrows1980problem,barrows1986taxonomy,wood2003problem,hmelo2004problem}. Unlike traditional teacher-centered methods, PBL integrates learning into meaningful problem scenarios, promoting active participation, collaboration, and the solving of complex real-life problems. Specifically, the PBL process typically involves presenting a problem, discussing it, conducting research and analysis, solving it, and summarizing the findings to gain knowledge and insights~\cite{barrows1980problem, boud2013challenge}.

\par Despite PBL's effectiveness in simulating real clinical diagnostic scenarios and engaging students in problem analysis and resolution, its primary application remains largely confined to medical teaching contexts~\cite{barrows1996problem,hmelo2004problem}. This limits the ability of students to engage in self-directed, targeted learning and improvement through PBL. Typically, the cases used in PBL are chosen by the instructors based on their experience and available clinical cases, which restricts the opportunity for targeted learning. Students often review case data selectively based on their experience and preferences, but without additional guidance, inexperienced learners may struggle to find specific cases of interest within a vast clinical dataset. Moreover, in clinical diagnostics, physicians are often required to conduct \textit{differential diagnosis}, which allows them to integrate and analyze multimodal data to construct a logical chain of reasoning~\cite{kassirer1991learning,norman2005research,elstein2002clinical}. As more evidence is gathered, the range of possible diagnoses narrows until a final diagnosis is determined. Senior physicians often rely on their existing knowledge to mentally organize this evidence. Although PBL aims to develop students' evidence-based reasoning through real-world scenarios, current practices often present data, such as medical records, laboratory results, and radiological images, in a fragmented manner with slides and paper formats, making it inconvenient to organize and analyze comprehensively. Online platforms like \textit{Lecturio}\footnote{https://www.lecturio.com/} and \textit{ eLfH}\footnote{https://www.wolterskluwer.com/en/solutions/uptodate} offer some integration for viewing and downloading materials, but still lack contextual reference information and interactive features. This limits the efficiency of specific analyses, such as reference range checking and image analysis, and hampers the integrated analysis of diverse data modalities. Consequently, this restricts the ability to construct and reason through logical chains in differential diagnosis based on comprehensive case information. Additionally, the methods for recording information during these analysis processes are often ineffective. Notes and electronic documents are frequently unstructured, making them difficult to review and complicating collaborative discussions. Online documents like \textit{Google Docs}\footnote{https://docs.google.com} also present challenges due to poor integration with original electronic medical records. These issues complicate the recording, review, and retrospective analysis, which are crucial for effective learning and improvement.


\par Existing research has explored various applications of PBL, including tools designed to enhance critical thinking~\cite{kek2011power,ali2019critical}, and the implementation and challenges of PBL in online courses~\cite{sistermans2020integrating,cheaney2005problem}. However, these studies generally extend PBL's characteristics without altering its presentation method, leaving students facing ongoing challenges in targeted training and learning. Simulation-based learning, which uses tools to recreate clinical scenarios, offers a way to advance PBL beyond traditional classroom environments~\cite{kneebone2005evaluating,gaba2004future,rosen2008history}. By leveraging technologies such as virtual reality (VR) and augmented reality (AR), these methods create immersive clinical settings~\cite{pottle2019virtual}, allowing students to engage in problem-solving with real-time interaction and feedback. Nevertheless, these studies often emphasize enhancing the realism of PBL simulations while overlooking the need for comprehensive analysis, summarization, and record-keeping essential to the learning process. Clinical Decision Support Systems (CDSS), developed using data-driven and AI techniques~\cite{cheng2021vbridge,choi2016retain,patel2019human,yang2019unremarkable,jacobs2021designing}, are also employed for analyzing and diagnosing case data~\cite{xie2020chexplain,calisto2023assertiveness,corti2024moving}. While CDSS aids physicians in clinical decision-making by detecting patterns and making predictions, it primarily focuses on data completeness and richness to enhance diagnostic accuracy and efficiency. However, it often overlooks the importance of forming a logical reasoning process. This contrasts significantly with the PBL approach, which emphasizes the development of metacognition and reasoning skills in educational contexts. \revision{Although existing tools such as \textit{NEJM Healer}~\cite{nejm_healer} provide case-based clinical reasoning support and training, they focus more on the reasoning and evaluation processes within the diagnostic workflow of specific cases. These tools lack considerations based on the PBL framework, such as actively exploring evidence through questioning (e.g. patient interviews) and explicitly presenting the logical reasoning process.}



\par Drawing on these insights, our goal is to gain a deeper understanding of students' logical and cognitive processes in clinical diagnosis to enhance their ability to apply knowledge and skills in real-world clinical settings. Following the general process of PBL in the medical field, which includes four key elements: identifying and defining the problem, generating hypotheses and learning issues, engaging in self-directed learning and research, and applying new knowledge to solve the problem~\cite{barrows1980problem,boud2013challenge}, we aim to develop a learner-centered PBL tool. This tool is designed to support students more effectively and facilitate the application of their knowledge and skills in clinical practice. To address existing research gaps and understand students' needs and barriers within the current PBL framework, we propose the following research questions \textbf{(RQ1-RQ2)}: \textbf{RQ1}: \textit{What are the needs and concerns of teachers and students during PBL practice and learning?} \textbf{RQ2}: \textit{Under what conditions, and in what ways, do students learn and improve through the PBL process?}

\par To answer these research questions, we conduct an iterative design process in collaboration with a team comprising HCI researchers, senior physicians (teachers), and novice physicians (students). These physicians perform as teachers and students in their medical education activities within the PBL framework. Our study consists of three phases: In \textbf{Phase One}, we collaborate with physicians to leverage their domain knowledge and gain a thorough understanding of current PBL practices in medical education and the underlying clinical expectations such as clinical reasoning and differential diagnosis. We examine the experiences of both students and teachers throughout the PBL process, focusing on how students learn and improve, the enhancements teachers expect, and the perceived benefits of their current training. We also investigate students' actual experiences and perceptions, the challenges they face during the process, and their needs and concerns. In \textbf{Phase Two}, we brainstorm learner-centered design features and potential scenarios that align with the PBL steps of hypothesizing, gathering information, analyzing, and concluding based on case studies. We iterate on these features through design cycles, ultimately determining the essential features needed to assist users in completing the PBL process and engaging in targeted learning for continuous improvement. In \textbf{Phase Three}, we recruit 13 physicians (both students and teachers) involved in the PBL process to be involved in a user feedback session to gather their feedback on these features and gain insights into how these features can promote a learner-centered PBL process. In summary, the contributions are as follows: 
\begin{itemize}
\item We collect perceptions of current PBL practices from both teachers and students, identifying their expectations, needs for improvement, and concerns related to PBL activities.
\item We identify the clinical reasoning-based differential diagnosis process, which involves evidence supplementation and further reasoning to reach a conclusion. Based on this, we designed features through iterations to \revision{develop a system} that supports learning and improvement throughout the PBL process.
\item We collaborate with physicians involved in the PBL process to gather feedback, gaining insights into how these features can aid learners during the PBL process.
\end{itemize}

\section{Background and Related Work}
\subsection{Problem-Based Learning in Clinical Education Practice}
\par Traditional medical education predominantly occurs in large classroom settings where teachers deliver instruction, and students passively receive information~\cite{murphy2021teacher}. This approach relies heavily on lectures and memorization, providing students with limited opportunities to engage and apply their knowledge to real-world problems. Consequently, there is often a disconnect between theoretical knowledge and practical application, hindering the development of students' active learning and problem-solving skills in real-life scenarios~\cite{barrows1996problem,boud2013challenge,reuell2019study}.

\par To enhance clinical practice in medical education, PBL is proposed as an educational approach that situates learning within the context of complex and meaningful problems~\cite{barrows1980problem,barrows1986taxonomy}. in PBL, students work collaboratively in small groups to solve practical problems, thereby acquiring embedded knowledge in these problems and developing problem-solving skills~\cite{barrows1986taxonomy,hmelo2004problem}. According to the PBL process proposed by Bound and Feletti~\cite{boud2013challenge}, students are presented with a problem, raise questions, discuss the problem, assign different questions to group members for research, reconvene to discuss and summarize new knowledge, and connect it to existing knowledge. Although specific instructional processes may vary among educators~\cite{barrows1980problem,boud2013challenge,schwartz2013problem}, they consistently involve problem identification, analysis, hypothesis formation, testing, and revision.

\par Research has shown that PBL significantly enhances clinical practice among learners, particularly in the development of critical thinking and self-directed learning skills~\cite{ozturk2008comparison,loyens2008self,ali2019critical}. By simulating real-world scenarios, PBL allows students to analyze and reason through clinical diagnostic processes. Building on the strengths of PBL, our work aims to bridge the gap between theory and practice in medical education. By focusing on learner-centered PBL practices, we seek to improve the training and development of students' clinical skills through these activities.



\subsection{Design Medical Diagnostic Tools with Physicians}
\par Existing research has detailed the process of collaborating with domain stakeholders for HCI design and development, highlighting the challenges involved, especially in the early stages or when stakeholders have limited experience in data science~\cite{buxton2010sketching,kross2021orienting,piorkowski2021ai}. Previous collaboration has primarily focused on designing user interfaces, with less emphasis on in-depth aspects such as overall problem formulation and model pipelines~\cite{delgado2021stakeholder,wang2023designing,yildirim2023investigating}. With the rise of human-centered design~\cite{beede2020human,amershi2019guidelines,li2023assessing,thieme2023designing}, research has begun to explore broader collaboration with domain stakeholders. The literature suggests that utilizing abstractions, sketches, and prototypes can significantly aid stakeholders in understanding relevant background knowledge, thereby enhancing communication and understanding~\cite{ayobi2023computational,yang2023harnessing,yildirim2024sketching}.

\par In the field of medical diagnostics, current efforts often involve extensive collaboration with physicians to understand their needs and design human-centered features to assist in the diagnostic process~\cite{xie2020chexplain,yang2023harnessing,calisto2023assertiveness,yildirim2024multimodal,hao2024advancing,sassmannshausen2024amplifying}. Designing in the medical field with stakeholders requires considering multiple factors, such as promoting shared decision-making (SDM) by involving both doctors and patients~\cite{hargraves2016shared,mangin2019think,patricio2020leveraging,hao2024advancing}, and enhancing communication between radiologists and clinical physicians to support radiology-based diagnoses~\cite{verma2021improving,xie2020chexplain,yildirim2024multimodal}. These collaborative design processes usually involve complex workflows and communication, highlighting the importance of humanistic medicine~\cite{chou2014attitudes,schattner2020essence}. For example, the clinical diagnostic process, especially for chronic diseases, involves disease detection, health status determination, communication with patients about treatment options, and finalizing shared decision-making results~\cite{hao2024advancing}. The radiology workflow involves physicians requesting image studies, radiologists performing scans and examinations, recording findings in reports, necessary communication between physicians and radiologists, and physicians making treatment decisions based on the reports~\cite{yildirim2024multimodal}. Furthermore, co-design with physicians often involves addressing their attitudes toward AI. A lack of knowledge and insufficient explanation may lead to mistrust in collaborating with AI~\cite{panigutti2022understanding,yang2023harnessing}, while it is equally crucial to avoid physicians becoming over-reliant on AI based on heuristics and shortcuts~\cite{bussone2015role, lee2023understanding}.

\par Recognizing the potential of PBL to enhance the application of knowledge in clinical settings, and addressing the limitation of current information presentation and organization methods for students in constructing differential diagnosis logic chains, we engaged in in-depth collaboration with physicians. Through this collaboration, we identified the existing clinical reasoning process and explored possible enhancements and features to facilitate student participation in this process during PBL activities. \revision{Existing studies show that frameworks such as User-Centered Design (UCD) enhance user satisfaction and learning outcomes in educational tool development~\cite{norman2013design}, while Activity Theory provides valuable insights for analyzing human-computer interactions and system design in educational contexts~\cite{engestrom2015learning}. Based on these insights, we conduct learner-centered design and consider users' interactions with features during the PBL process. By incorporating the principles of PBL, we encourage students to drive their exploration and analysis of evidence through questioning, while explicitly updating the logical chain throughout the analytical process. This distinguishes our approach from tools that primarily focus on clinical reasoning through case data analysis~\cite{nejm_healer}}.


\subsection{Human-Centered Medical AI}

\par With the development and application of AI, researchers increasingly emphasize the human aspect in AI collaboration workflows, specifically focusing on human-centered AI research~\cite{xie2020chexplain,thieme2023designing,yildirim2023creating}. 

\par In the medical field, human-centered medical AI research encompasses the study of different stakeholders involved in the medical process. Specifically, this research explores physicians' collaboration in clinical diagnosis decision-making with AI~\cite{calisto2021introduction,calisto2022breastscreening,cheng2021vbridge,ouyang2023leveraging}. This includes utilizing AI assistance in classifying and explaining medical imaging diagnosis in breast cancer~\cite{calisto2021introduction}, incorporating ML explanations into physicians' decision-making workflows~\cite{cheng2021vbridge}, and considering patient characteristics like history and treatment preferences to provide diagnostic recommendations~\cite{cai2019hello,mckinney2020international,bulten2022artificial}.
Some studies focus on enhancing collaboration among physicians based on diagnostic processes~\cite{verma2023rethinking,gu2021lessons,xie2020chexplain,yildirim2024multimodal}. Xie et al.~\cite{xie2020chexplain} clarified the understanding and needs of physicians and radiologists regarding AI-enabled medical analysis and interaction. Subsequent research further explored the radiology workflow, investigating the needs and expectations of radiologists and clinical physicians concerning AI-assisted radiology workflows~\cite{yildirim2024multimodal}. Beyond collaboration among physicians, research also aims to improve communication and decision-making between physicians and patients~\cite{ankolekar2019development,seljelid2022feasibility,hao2024advancing}. Studies on patient-centered decision-making~\cite{mangin2019think,patricio2020leveraging} demonstrate that active patient participation in clinical decisions leads to higher satisfaction and better treatment outcomes~\cite{kuosmanen2021patient}. The integration of AI into shared decision-making has been explored, such as developing tools to assist patients in making decisions about prostate cancer treatment~\cite{ankolekar2019development} and facilitating clinical decision-making for older cancer patients~\cite{hao2024advancing}.


\par While human-centered AI for stakeholders in medical diagnosis has been widely researched, its application in medical education remains relatively unexplored. \revision{Despite current tools and platforms~\cite{nejm_healer,uptodate,osmosis} being utilized for the learning process in medical education, these tools primarily focus on decision-making in clinical practice or general educational learning, leveraging AI to analyze and assess users' learning performance and make recommendations.} Building on human-centered AI concepts and research in the medical field, our study focuses on learner experiences in PBL activities within medical education. We aim to design learner-centered features that \revision{leverage interactive elements and AI to facilitate evidence exploration and analysis, enhancing students' PBL processes and improving their clinical reasoning skills and diagnostic practice, bridging the gap between knowledge and practical application}.

\section{Overview of Clinical Reasoning-based Differential Diagnosis}
\par
\begin{figure*}[h]
\centering
\includegraphics[width=\textwidth]{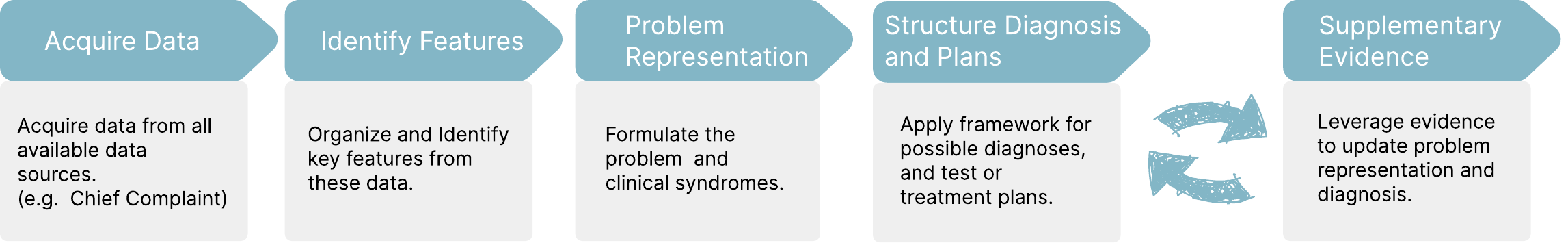}
\caption{Overview of Differential Diagnosis workflow. The process consists of acquiring data, identifying key features from data, formulating problem representation from these features, developing diagnosis in a structured framework with the following plans for test and treatment, and iteratively updating diagnosis and plans with supplement evidence from test results.}
\label{fig:workflow}
\end{figure*}

 Differential diagnosis is a systematic approach employed by physicians to determine the disease or condition responsible for a patient's symptoms~\cite{kassirer1989diagnostic,swenson1999differential}. As demonstrated in \autoref{fig:workflow}, this process starts with data collection, including patient history and physical examination, to gather initial information. Physicians then organize this data to identify key features relevant to clinical diagnosis. During the differential diagnosis process, physicians use these features to formulate a problem representation: a concise summary using precise medical terminology that highlights the most relevant aspects of the patient's condition. \revision{This representation includes semantic qualifiers that frame the patient's symptoms in familiar terms, facilitating the integration of related findings into clinical syndromes.} Finally, physicians apply a framework (e.g. anatomic or physiological) to generate a structured list of potential diagnoses.

\par For each potential diagnosis, physicians evaluate the likelihood of each condition based on the clinical presentation and the patient's background, including factors such as age and occupation. During the differential diagnosis process, physicians develop appropriate diagnostic test plans. The results of these tests are used to support or rule out potential diagnoses from the list. As shown in \autoref{fig:workflow}, with the test results providing additional evidence, physicians integrate and synthesize the collected data, comparing clinical findings with known disease patterns and pathophysiology. This allows them to update their problem representation and diagnostic reasoning, leading to revised test or treatment plans. This iterative process incorporates new evidence and continuously refines the diagnostic reasoning. Through critical thinking and logical reasoning, physicians narrow down the differential diagnostic list to identify the most likely diagnosis. The process of confirming a diagnosis resembles a funnel-shaped curve, where the scope narrows as evidence accumulates, ultimately leading to a final conclusion~\cite{blois1980clinical,gu2021lessons}. \revision{Diagnostic approaches can be categorized into two types based on cognitive strategies: \textit{heuristic}~\cite{mcdonald1996medical,marewski2012heuristic} and \textit{metacognitive}~\cite{cloude2022role,merkebu2024case}. \textit{Heuristic} reasoning relying on past experiences and intuition to identify disease patterns quickly, while \textit{metacognition} involves reflecting on and adjusting diagnostic decisions based on new information. In differential diagnosis, both \textit{heuristic} reasoning and \textit{metacognition} are used to identify potential diagnoses and refine decisions throughout the process~\cite{packer2019approaches}.}

\par PBL aims to train students to solve real-world problems in complex, practical contexts, similar to the process of differential diagnosis, by employing systematic thinking and evidence-based support \revision{clinical reasoning}. This approach involves heuristic reasoning to generate initial hypotheses and metacognition to reflect on and adjust thinking processes. While much of the research in medical education focuses on case-based learning beyond clinical decision-making~\cite{sultanum2018more,sultanum2022chartwalk,ouyang2023leveraging}, there is relatively little research that considers design from the perspective of logical reasoning-based on differential diagnosis. Our study addresses these considerations by exploring design spaces that assist medical students in enhancing their \revision{clinical diagnostic reasoning through the PBL process}.

\section{Method}
\par Our study is structured into three phases. In \textbf{Phase One}, we conduct discussion sessions with senior and novice physicians involved in the current PBL process. We specifically gather insights from their roles as teachers (senior physicians) and students (novice physicians), identifying needs and barriers from each perspective, and analyzing the findings. In \textbf{Phase Two}, we brainstorm learner-centered designs and scenarios, and iterate on these designs to develop features that effectively support students in completing the PBL process, with a focus on clinical reasoning-based differential diagnosis. In \textbf{Phase Three}, we propose a prototype for a learner-centered PBL process. We then conduct evaluations with participants involved in PBL activities to gather qualitative and quantitative feedback on the usability of the system features. All studies were conducted with the approval of the Institutional Review Board (IRB) and with informed consent from the participants.

\section{Phase One: Identify Current Practices and Challenges in PBL}
\subsection{Procedures and Data Analysis}
\par To answer \textbf{RQ1} and \textbf{RQ2}, we conducted discussion sessions in \textbf{Phase One} with physicians (they are not co-authors of this work) from our collaboration team. This includes senior physicians serving as teachers in PBL activities: \revision{a Neurosurgery Attending physician S1 (female, age:36) and an Orthopedic Associate Chief physician S2 (male, age: 47), and novice physicians participating as students: two Orthopedic Intern physicians N1 (male, age:26), N2 (male, age:27).}
We held four separate discussion sessions with each physician, each session involved semi-structured interviews featuring open-ended questions.

\par With the senior physicians, we explored their typical organization of PBL teaching activities, covering aspects such as case selection, process explanation and guidance, and post-learning summarization. We also delved into their expectations for student learning during the PBL process, asking questions like ``\textit{Which aspects do you think PBL can enhance in students?}'' and ``\textit{What are the similarities and differences between PBL and clinical case reasoning, and how can PBL promote students' clinical reasoning skills?}'' Additionally, we gathered their insights and concerns on the current process.

\par For the novice physicians, we focused on their perspectives regarding learning and improvement during the PBL process. We posed questions like ``\textit{How do you think you can achieve improvement in the current PBL process?}'' and ``\textit{How does the current PBL process help you acquire clinical diagnostic skills?}'' We also collected their views on PBL, including their concerns and needs in the current practice, with questions like ``\textit{Do you think the current information presentation and analysis process in PBL meets your learning needs, and what areas can be improved?}''.

\par We collected data by recording the interviews with the consent of the interviewees. After transcribing the recordings, we applied thematic analysis~\cite{braun2006using,braun2012thematic} to analyze the data. Two researchers from the study team conducted the initial coding, generating labels based on the initial codes. Using an affinity diagram~\cite{corbin2014basics}, we grouped these labels and organized them into themes with corresponding definitions. Following \textbf{RQ1} and \textbf{RQ2}, which focus on perceptions and improvement in PBL, we organized the findings into two themes: \textbf{Perception of the PBL Process} and \textbf{Acquisition of Learning and Improvement}.

\subsection{Perception of PBL Organization and Implementation}
\subsubsection{Teachers' case selection reflects their expectations for students to grasp the relevant knowledge}
\par Senior physicians, serving as teachers in medical education activities, are responsible for selecting cases for PBL. Their case selection typically considers various factors, including alignment with the curriculum, the establishment of a learning gradient, and relevance to clinical scenarios.

\par PBL is typically implemented in classroom settings to enhance students' understanding and application of knowledge in real-world scenarios. Therefore, cases are often designed around specialized content that aligns with current diagnostic practices within specific disciplines. In designing these cases, senior physicians emphasize the importance of balancing complexity and familiarity. They recognize that medical learning can be categorized into two primary forms: reinforcing existing knowledge and acquiring new knowledge to expand expertise. As one senior physician (S1) noted, \textit{``For teaching, we usually start with classic cases to build a strong foundation, and then we mix [in] some complex cases to gradually increase the difficulty and help [develop] their problem-solving skills.''} Physicians acknowledge that in teaching scenarios, PBL primarily aims to ensure mastery of fundamental concepts and common conditions, consistent with the philosophy of Occam's Razor in differential diagnosis~\cite{packer2019approaches} which prefers the simplest explanation and the most common conditions over rare ones, to reduce unnecessary complexity and avoid overdiagnosis. As another senior physician (S2) remarked, \textit{``We mainly design cases around classic examples because we've noticed that if we introduce too many complex [cases], students tend to jump to rare conditions when they see similar symptoms.''}

\par To promote the practice of clinical scenarios, physicians often integrate elements of real-world humanistic medicine into case presentations. As S2 noted, \textit{``Even though we mainly assess students' knowledge, we still present case descriptions with medical data in a storytelling format. We include elements of humanistic medicine, like communication during clinical decision-making and [ensuring] patient emotional stability, to help train students' overall skills in real clinical settings.''}

\subsubsection{The teacher's role as a guide facilitates student discussion and encourages active exploration}
\par In PBL activities, students take an active role, while teachers focus on organizing and guiding the learning process. Teachers typically introduce relevant research questions related to the cases to foster in-depth thinking and discussion among students. As S1 explained, \textit{``We usually start by sharing information about the patient's chief complaint, their medical history, and their current illness, along with their background and how their symptoms began. At this point, students need to make initial judgments based on the symptoms presented. We encourage their thinking and help guide their reasoning by posing questions and providing evidence throughout this process.''}

\par Senior physicians point out that PBL cases are typically presented in a hierarchical structure, mirroring the differential diagnosis process. After initial analysis, students receive additional evidence and engage in further analysis tasks, which helps sustain their engagement and focus. As S2 explained, \textit{``[...In the later stages,] when students get more information like radiology images and Complete Blood Count (CBC) results, they need to do a thorough analysis and reasoning to reach a diagnosis. I think this step-by-step approach keeps students engaged and interested, and it definitely helps them apply and master their knowledge better than traditional methods.''}

\subsubsection{Students acknowledge that PBL increases their engagement and motivation, and they express a desire for more integrated information organization}
\par In line with teachers' expectations for student knowledge acquisition, students recognized the benefits of PBL in promoting their participation. As N1 noted, ``\textit{PBL really gets us more engaged in learning. Just sitting [there] passively listening to lectures can get boring and tiring, but when we're involved in exploratory activities, it boosts our motivation and participation. Our thinking becomes more active and efficient during discussions.}''

\par Students also emphasized the need for integrated organization of diagnostic data to support their analysis and discussions, rather than relying on existing CDSS tools that merely present results. As one student N1 explained, \textit{``I get that the whole point of PBL is to help us learn through analysis, like discussions. So, we're more focused on the analysis process [itself], not just on seeing the results. We need to connect and present different data to improve how we retrieve information and do hands-on analysis.''}  Besides the diagnostic data, the organization of information during the analysis process is also crucial. Students expressed that properly structuring the analyzed information could significantly enhance their ability to analyze and discuss cases. One student (N2) noted, ``\textit{During our discussions, we often need to jot down and organize our thoughts to help [with] the analysis. For example, When we get a patient's symptoms, we list possible causes and then make a differential diagnosis as more clinical data comes [in]. Having everything structured really helps our analysis and makes us more efficient.}''

\subsection{The Acquisition of Learning and Improvement}
\subsubsection{Teachers strive to impart a wide range of skills and improvements to their students, but they are often constrained by limited classroom time and the need to address the broader requirements of the majority}
\par Senior physicians, serving as teachers in medical education, express a strong desire to share as much of their accumulated experience with students as possible. However, they face constraints due to limited classroom time, which necessitates compromises. As one physician (S2) explained, ``\textit{There are so many classifications and variations of diseases within the same specialty. For example, arthritis alone has over 100 [types], with osteoarthritis and rheumatoid arthritis being the most common. With the limited [classroom] time, we cannot cover and practice every single type, so we have to be selective about the cases we choose.}''

\par However, they also acknowledge that these selections are often guided by intuitive heuristic experience rather than systematic empirical evidence. As one physician (\revision{S2}) noted, ``\textit{Currently, we do not have tools to help us systematically screen cases based on [things] like incidence rates and clinical manifestations. We usually just pick what we think are the important and classic clinical cases.}'' This selection process may involve cognitive biases that could influence the final judgment and choice.

\par Additionally, senior physicians recognize that the teaching context often necessitates prioritizing the needs of the majority, which can result in limited individualized instructional support. As S1 noted, ``\textit{In classroom teaching, we always have to think about the needs of the majority, like whether most students have mastered the planned knowledge through the cases. Unfortunately, that means we don't give [enough] attention to individual [needs].}'' This approach can leave certain student needs unmet, such as personalized supplementary training and opportunities for advanced skill enhancement.

\subsubsection{Learning that is oriented toward clinical practice is effective, but the process of constructing clinical reasoning skills through PBL remains implicit and subtle}
\par Both teachers and students acknowledge PBL as an effective approach for integrating learned knowledge with clinical practice. Teachers view it as a valuable way to pass on their accumulated experience to students. As S2 noted, ``\textit{Throughout the PBL process, we keep presenting evidence alongside the case introductions, which really helps illustrate the core of clinical diagnosis and treatment. This way, students can learn and improve through hands-on experience, which is much more effective than just telling them our conclusions.}''

\par They also recognize that the construction of clinical reasoning and logic in differential diagnosis is often implicit and subtle, a challenge mirrored in the PBL learning process. ``\textit{We might not explicitly guide students in constructing logical frameworks, like writing on the board or using other methods to make them aware of the reasoning and logical chains at play. This aspect still leaves room for students to explore. (S2)}'' Students' feedback is consistent with this understanding: ``\textit{PBL activities help [improve] my clinical practice skills, but sometimes the process feels a bit inefficient. During the analysis and discussion, there can be multiple speculations or treatment suggestions for the same condition, and while discussions do refine and clarify things, the logical reasoning behind them isn't always clear. It often stays in our minds and gets expressed verbally. We might use sketches or mind maps to help, but there isn't enough support or guidance for that (N2)}.'' This highlights the need for more structured approaches in executing the clinical reasoning process.

\subsubsection{Student express a desire for a more manageable and accessible PBL learning process}
\par Students identified chances for improvement in their learning experience. Specifically, they expressed a desire for a more participatory and manageable learning process, including the ability to select targeted cases and establish more effective feedback mechanisms.

\par As noted, teachers typically select cases to help students master classic cases and basic knowledge, but this approach may not meet all students' needs. As N2 observed, \textit{``I understand that the textbook symptoms are important, but I also know that in real clinical practice, diseases can present with different subtypes and manifestations. So, systematic learning should focus on identifying the gaps between my knowledge and practical scenarios and conducting targeted analysis and training.''}  In this context, Hickam's Dictum~\cite{borden2013hickam}, which counters Occam's Razor, is relevant. It emphasizes the importance of thorough analysis and evaluation rather than prematurely seeking simple correlations.

\par Senior physicians also emphasized the crucial role of reflection and review in the PBL process. S1 noted, \textit{``Reflection and review are really key parts of the learning process. From experience, we know that learning from our mistakes is a powerful way to improve ourselves.''}  However, feedback has highlighted some shortcomings in the current mechanism. As noted by N1, \textit{``Right now, teachers typically give final summaries and comments after PBL [activities], but these are often more improvised than systematic. Even though we submit analysis reports, they usually don't follow a standardized format for reviewing the activity.''}  This absence of a standardized approach complicates subsequent reviews, making it challenging to trace specific content and insights, and diminishing the initiative for thorough retrospection and reflection.

\section{Phase Two: Deriving Design Space and Solutions for Enhancement}
\subsection{From Findings to Design Strategies}
\par Based on feedback from physicians, we analyze these findings to identify their key concerns, categorizing them into two themes: \textbf{What to Learn} and \textbf{How to Learn and Improve}. These themes are further summarized as \textit{Content Setting} and \textit{Analysis Process}.

\par \textit{Content Setting} involves the selection of cases and data in PBL activities, focusing on factors such as choosing cases that align with the teaching material, establishing a progression in difficulty from simple to complex, covering key points within a limited timeframe, and balancing individual specific needs with the broader group requirements. \textit{Analysis Process} consists of two aspects of PBL analysis: procedural organization and information organization. Procedural organization addresses the steps and execution of clinical reasoning and differential diagnosis during PBL analysis. Information organization involves the presentation, organization, and recording of information throughout the analysis process, including case details, emerging ideas and hypotheses, and insights and reflections. The codebook for these findings is presented in \autoref{tab:codebook}.

\begin{table*}
\caption{The codebook for findings derived from discussion sessions.}
\label{tab:codebook}
\begin{tabular}{p{3cm}p{3cm}p{4cm}p{4cm}}
\hline
\textbf{Theme} & \textbf{Code} & \textbf{Sub-Code} & \textbf{Definition} \\ \hline

\textbf{Content Setting} & \textbf{Case Selection} & \textbf{Related to instructional knowledge} & Instructors select relevant cases for PBL based on the instructional content to help students reinforce and enhance their understanding. \\ \cline{3-4}
& & \textbf{Gradual progression of difficulty/complexity levels} & The learning process is gradual, with appropriate expansion upon mastering the foundational content. \\ \cline{3-4}
& & \textbf{Adaptation to individual learner needs} & Learners have specific needs for personalized training and development during the PBL process. \\ \hline

\textbf{Analysis Process} & \textbf{Procedural Organization} & \textbf{Logical Construction during Diagnostic Reasoning} & The diagnostic reasoning process involves structured development of logical chains. \\ \cline{3-4}
& & \textbf{Evidence-based Reasoning and Differential Diagnosis} & The reasoning and differential diagnosis are conducted based on evidence. \\ \hline

 & \textbf{Information Organization} & \textbf{Consistent Presentation of Information} & The case information needs to be presented consistently for contextual analysis and viewing. \\ \cline{3-4}
& & \textbf{Recording and review of the Analysis Process} & Relevant information from the analysis process, such as notes and annotations need to be organized and recorded to help review. \\ \hline
\end{tabular}
\end{table*}

\par Building on the summarized differential diagnosis workflow (\autoref{fig:workflow}), we focus on the application of clinical reasoning in differential analysis throughout the PBL process. This includes the presentation, analysis, and organization of information and analysis. The process involves clinical information such as the patient's chief complaints, physical examination findings, test results, and other diagnostic evidence, as well as analyzed data such as sketches, mind maps, annotations, and insights generated during the analysis. Proper presentation and organization of this information facilitate analytical discussions and enhance learning outcomes from PBL activities. Based on this analysis, we structured the entire PBL process into three stages: \textit{Data Construction}, \textit{Information Analysis}, and \textit{Record and Review}. We then categorized the codes within these themes according to these stages, which informed the design features for each stage.


\subsection{Design and Development of e-MedLearn}
\par
\begin{figure*}[h]
\centering
\includegraphics[width=\textwidth]{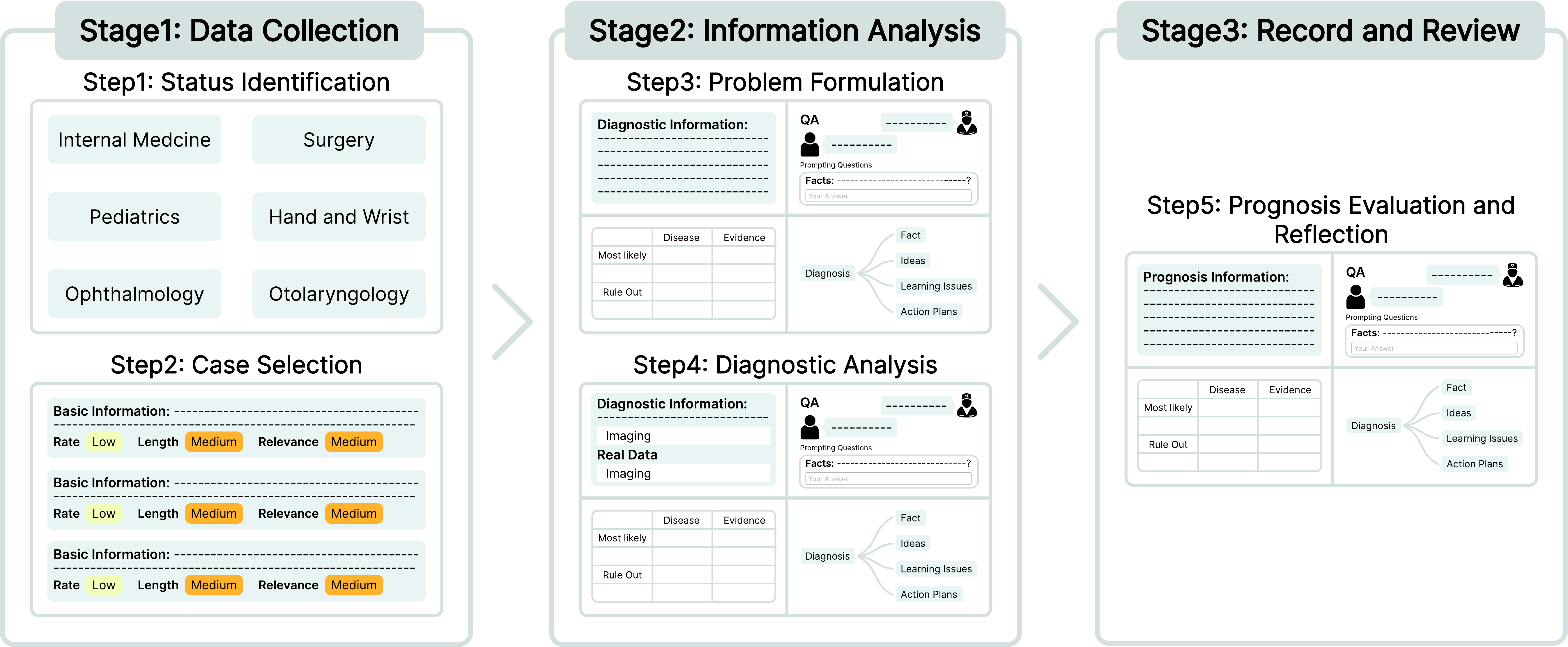}
\caption{The stages and steps of the \textit{e-MedLearn} prototype system pipeline.}
\label{fig:pipeline}
\end{figure*}

 \revision{In the design of \textit{e-MedLearn}, we adopted the UCD framework~\cite{norman2013design}, conducting multiple rounds of interviews with medical students and teachers in our collaboration team to ensure that the system's design and features meet user needs. Throughout the design process, various ideas were explored and iteratively refined based on user feedback. Initial prototypes integrated the entire case analysis process into a single interface for analysis (\autoref{fig:prototype}). However, user feedback revealed that this design created a complex interface, and despite through the navigation structure, it still imposed a significant cognitive load. Following further discussions, we decided to streamline the design by dividing the features into distinct steps, informed by the results of the thematic analysis. By reorganizing the content from \autoref{tab:codebook}, we developed a five-step presentation process designed to enhance students' learning and improvement throughout the PBL process. As shown in \autoref{fig:pipeline}, these steps are aligned with the three stages of the PBL process: \textit{Data Construction}, \textit{Information Analysis}, and \textit{Record and Review}. This structured approach ensures that students receive comprehensive support as they progress through the entire PBL learning process. A step-by-step demonstration of the system interface is provided in \autoref{fig:walkthrough}.}

\begin{figure*}[h]
\centering
\includegraphics[width=\textwidth]{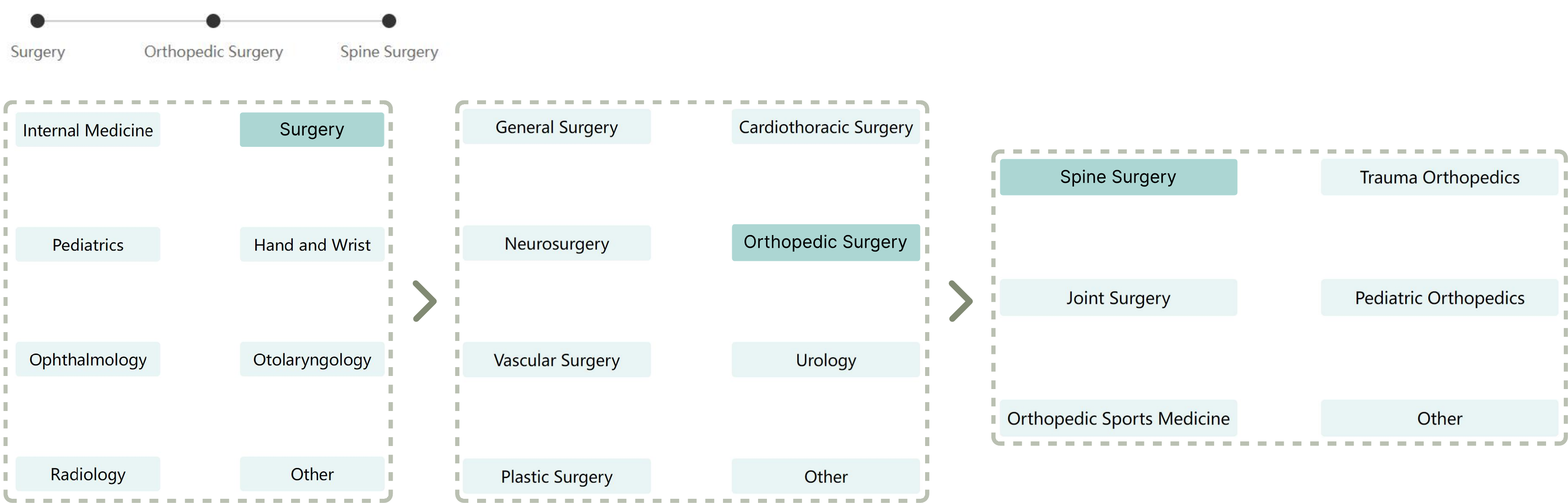}
\caption{In Step 1: Status Identification, the \textit{e-MedLearn} platform provides a top-down, multi-level selection process to help users determine the content of their learning.}
\label{fig:Step1}
\end{figure*}

\begin{figure*}[h]
\centering
\includegraphics[width=\textwidth]{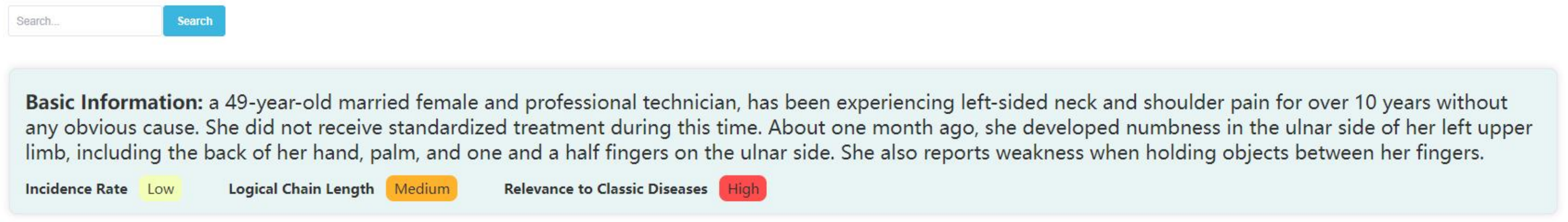}
\caption{In Step 2: Case Selection, relevant cases are presented as cards, with each card displaying difficulty scores for reference. Users can also search cases by content.}
\label{fig:Step2}
\end{figure*}

\subsubsection{Stage One: Data Construction}
\par In \textbf{Stage One}, users prepare for analysis based on the preliminary review of collected data. In \textit{Step 1: Status Identification} (\autoref{fig:Step1}), users assess their learning status by determining relevant knowledge areas, such as specialties and types of diseases. We adopt a multi-level classification mechanism that helps users identify their current learning status by navigating from general categories to more specific ones (e.g. Surgery\raisebox{-0.6ex}{\includegraphics[height=2.5ex]{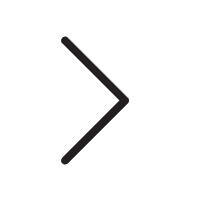}}Orthopedic Surgery\raisebox{-0.6ex}{\includegraphics[height=2.5ex]{Img/Arrow.png}}Spine Surgery), which aids in selecting appropriate cases for further study. In \textit{Step 2: Case Selection} (\autoref{fig:Step2}), the system supports targeted case selection for PBL activities. Users, guided by their understanding of their learning status, choose relevant cases with system assistance. The interface presents cases categorized under the current classification along with brief descriptions. Cases were initially evaluated by two senior physicians in our team, who evaluated difficulty levels based on three dimensions: incidence rate, length of logical chain, and relevance to classic diseases (the detailed criteria are shown in the Appendix) for reference. Users can also use the search bar to find cases by matching keywords in the descriptions.

\begin{figure*}[h]
\centering
\includegraphics[width=\textwidth]{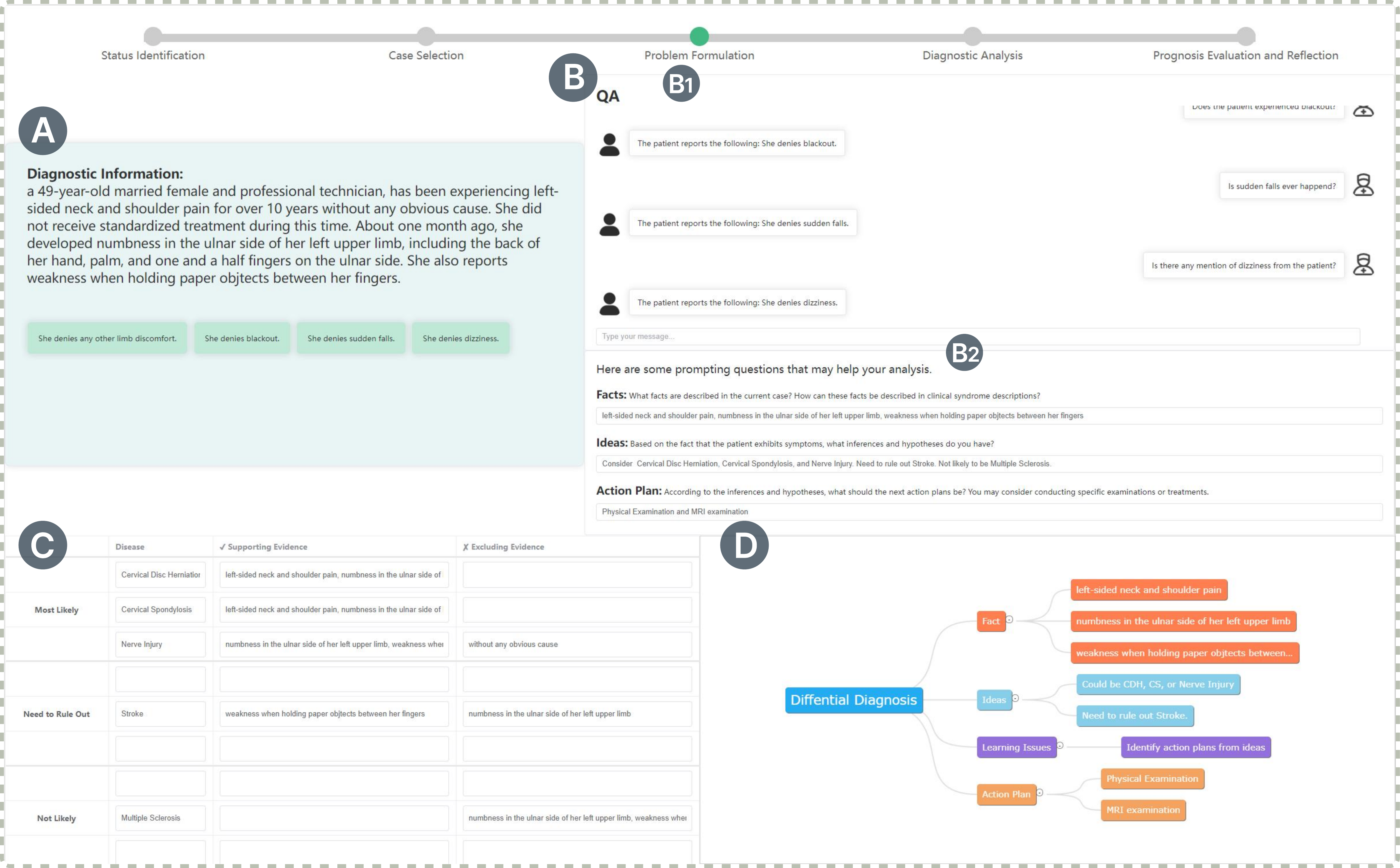}
\caption{Step 3: The Problem Formulation interface includes: (A) Case information card. (B) The prompting QA section. (C) The template mind map to organize and record analyzed information.}
\label{fig:Step3}
\end{figure*}

\subsubsection{Stage Two: Information Analysis}
\par In \textbf{Stage Two}, the system provides explicit support for analyzing and organizing information based on the clinical reasoning process of differential diagnosis. Leveraging GPT-3.5-turbo-16k model~\cite{brown2020language}, we have integrated AI-driven QA features to facilitate information analysis. \revision{Specifically, we extracted data from each case and input it into the model. Using specifically designed structured prompts, the model generates responses to user inquiries about case details, enabling a more thorough and thoughtful exploration during analysis. The detailed prompt template is present in \autoref{tab:structured_prompt}.} Collaborating with physicians, we developed a set of predefined template questions across various steps (as shown in the Appendix) following the FILA (Facts, Ideas, Learning Issues, Action Plans) framework~\cite{choo2012scaffolding}. These questions are designed to prompt users in case analysis and reasoning.

\par In \textit{Step 3: Problem Formulation} (\autoref{fig:Step3}), users are guided to analyze and organize the preliminary case information, identify key ideas, and formulate subsequent action plans. This step focuses on helping users construct and evaluate logical chains based on initial evidence. The interface consists of (A) Preliminary case information; (B) A QA section with prompting questions; (C) A diagnostic list for reasoning updated with evidence; and (D) A FILA-based mind map template for analysis organization and record. The diagnosis list and mind map content are consistently shared across different steps to ensure accurate information integration and recording throughout the analysis process. We deliberately processed the case information to omit descriptions of ``denial'' symptoms, such as the denial of fever, within the case study. These unexpressed symptoms, termed as ``\textit{negative symptoms}'', are identified through the physician's proactive questioning during diagnosis. They are crucial in reflecting the thoroughness of the physician's diagnostic reasoning. Users can obtain this information by asking relevant questions in the QA section, enabling them to practice their questioning and reasoning skills during the diagnostic process. When asked about negative symptoms, \revision{the model is prompted to generate responses based on the patient's actual condition. If the case information includes relevant details, the system will display them accordingly. Otherwise, it will indicate that the information is not relevant to the analysis.} In the diagnosis list, users write down their initial thoughts on the case, we design the list into three categories: \textit{Most likely}, \textit{Need to Rule Out}, and \textit{Not Likely}. This design addresses the importance of evaluating conditions by likelihood, ensuring that critical diseases are not missed, while also considering the dynamic nature of diagnoses, where diseases may shift from improbable to probable. In the QA section, these prompting questions guide users' analysis. The responses to these questions are then integrated under the corresponding factor (Facts, Ideas, Learning Issues, Action Plans) in the mind map, where users can further edit and refine the information. The prompting questions cover: A) \textbf{Facts}: ``\textit{What facts are described in the current case? How can these facts be contextualized in clinical syndrome descriptions?}''; (B) \textbf{Ideas}: ``\textit{Given that the patient exhibits <symptoms>, what inferences and hypotheses do you propose?}''; and C) \textbf{Action Plans}: ``\textit{Based on these <inferences and hypotheses>, what should the next action plans be, such as conducting specific tests or treatments?}'' Additionally, users can proactively inquire about patient information not mentioned in the dialogue area such as surgical prognosis or allergy history to further support the differential diagnosis process.

\begin{figure*}[h]
\centering
\includegraphics[width=\textwidth]{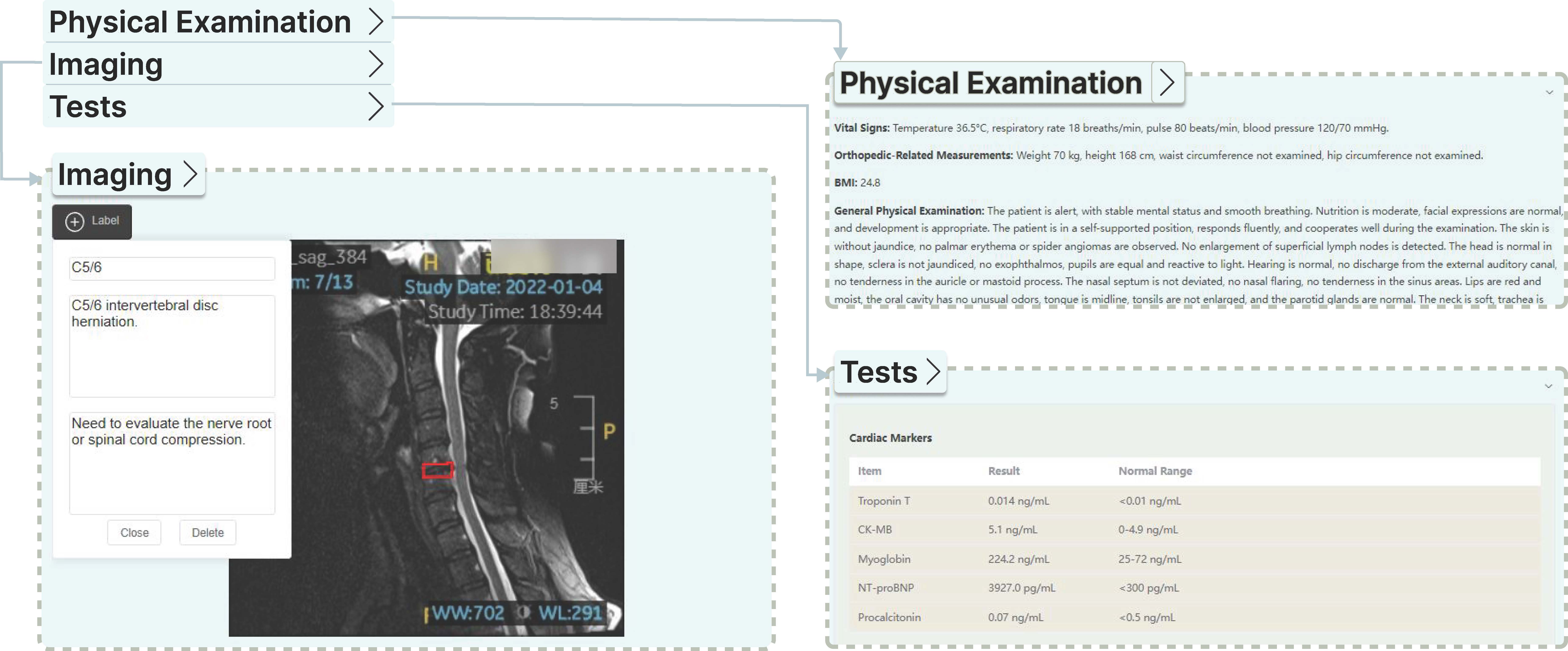}
\caption{In Step 4: Diagnostic Analysis, the case information is presented in a collapsible section, including imaging (images), physical examination (text), and test results (indicator).}
\label{fig:Information panel}
\end{figure*}

In \textit{Step 4: Diagnosis Analysis}, users perform diagnostic analysis using data from additional examinations or tests. This involves updating their initial problem formulation and refining their reasoning to plan subsequent tests or treatments. The system organizes and presents further case scenarios and narratives, through collapsible sections (\autoref{fig:Information panel}), displaying various types of information, such as physical examination results (text), radiology images (image), and test results (indicator). For imaging data, the system includes annotation features that allow users to draw, label, and add notes to specific areas for marking. Test results are presented in a clear format, listing ``item, result, normal range'' for easy viewing and comparison. The QA section offers prompting questions to guide users through analyzing the results and making decisions. Example questions include: A) \textbf{Facts}: ``\textit{What results does the <test> show, and what do they indicate?}''; B) \textbf{Ideas}: ``\textit{What inferences can be drawn from the <findings>, and how do they assist in further differentiation and discovery?}''; C) \textbf{Learning Issues}: ``\textit{Are the <descriptions> of the patient consistent with your problem representation, and do the <tests> address previous considerations?}''; and D) \textbf{Action Plans}: ``\textit{Based on this analysis, what additional tests or treatment plans should be considered?}'' Following the prompt, users can inquire about the tests and examinations, with the results displayed on cards for each test type, facilitating comparison with subsequent test data presented in the collapsible sections. Users update their thoughts and problem representation in the mind map provided during analysis. This iterative process, indicated by the updated data (\autoref{fig:workflow}-\raisebox{-0.6ex}{\includegraphics[height=3ex]{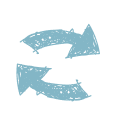}}), continues until the differential diagnosis is complete. At the end of this process, users consolidate their reasoning based on the evidence to finalize the diagnosis and treatment plan.

\subsubsection{Stage Three: Record and Review}
\par During \textbf{Stage Three}, users engage in the discussion and evaluation of the prognosis, organizing and recording their analysis and insights for further review and reflection. The interface, similar to \autoref{fig:Step3} but with updated information, is detailed in the Appendix.

\par In \textit{Step 5: Prognosis Evaluation and Reflection}, users examine the case's details regarding diagnosis, treatment, and prognosis, including diagnostic results, treatment methods, and recovery outcomes, with consideration of medical humanities. The QA interface provides prompting questions designed to facilitate both learning about the prognosis and guiding users through the review and reflection of their diagnostic process based on the case narrative. For prognosis, the prompting questions include: A) \textbf{Facts}: ``\textit{What key prognosis details are highlighted in the case?}''; B) \textbf{Learning Issues}: ``\textit{What aspects, such as patient-family communication or social considerations in treatment decisions, were previously overlooked and needs attention?}'' For reviewing and reflecting on the analysis, the prompting questions address: A) \textbf{Facts}: ``\textit{Are the facts related to diagnosis and treatment consistent with your previous analysis? What discrepancies or overlooked areas are present?}'' B) \textbf{Ideas}: ``\textit{What caused any inconsistencies or oversights? Where did the issues arise in the mind map's analytical pathway?}'' C) \textbf{Learning Issues}: ``\textit{What lessons can be drawn from this case analysis? What areas of improvement were identified, and which areas need further enhancement?}'' Users use these prompts to review their analysis process, identify errors or deficiencies through mind maps and notes, and reflect on their diagnostic thought processes. With traceable and accessible case data and analysis notes, users can systematically record insights, reflections, and summaries. Additionally, they can export these notes and summaries for further review and retrospection.


\section{Phase Three: Stimulating User Feedback}
\par In this phase, we engaged physicians experienced in PBL to participate in feedback sessions. The experimental design was centered on the system's goal of enhancing students' clinical reasoning, knowledge acquisition, and overall improvement within the PBL process. \revision{Instead of replacing traditional PBL education in classroom settings, our approach focuses on the advancement of clinical reasoning within the PBL framework to facilitate targeted learning and improvement.}


\par \revision{We conducted a two-stage study to evaluate the effectiveness and usability of \textit{e-MedLearn}, gathering both quantitative and qualitative data. First, a controlled study was carried out to evaluate the impact of integrating clinical reasoning into the PBL process on improving differential diagnosis. Second, a series of testing interviews focused on the PBL process and user feedback were conducted. This approach provided insights from participants, helping us identify the system's strengths, areas for improvement, and its overall effectiveness and usability in advancing PBL within medical education.}



\subsection{Controlled Study}
\subsubsection{Participants}
\par \revision{We recruited $19$ participants ($11$ males, $8$ females, average age: $27.5$) through mailing lists and snowball sampling within a local hospital, facilitated by our collaborating team. The participants are not co-authors of this work and did not participate in the discussion sessions during \textbf{Phase One}. They are novice physicians with prior PBL experience as part of their continuous medical education. Detailed demographic information about the participants (S1-S9) is provided in \autoref{tab:between_students}. We also involved two senior physicians ($2$ males, average age: $42$) as teachers in the PBL process to evaluate the participants' performance in the study. The demographic information of the senior physicians (T1, T2) is provided in \autoref{tab:between_teachers}.}

\begin{table}[h!]
\caption{Demographic information of participants in the controlled study: PBL participation encompasses various levels of engagement, including regular involvement (Routine) and participation based on individual needs (On-demand).}
\label{tab:between_students}
\begin{tabular}{p{0.2cm}p{0.8cm}p{0.3cm}p{1cm}p{0.8cm}p{1.6cm}p{1.5cm}}
\hline
\textbf{ID} & \textbf{Gender} & \textbf{Age} & \textbf{Title} & \textbf{PBL Role} & \textbf{PBL Participation} & \textbf{Clinical Domain} \\
\hline
S1  & Male   & 26 & Resident & Student & Routine & Orthopedics \\
S2  & Male   & 24 & Intern & Student & Routine & Orthopedics \\
S3  & Male   & 26 & Resident & Student & Routine & Orthopedics \\
S4  & Male   & 28 & Resident & Student & Routine & Orthopedics \\
S5  & Female & 27 & Resident & Student & On-demand   & Orthopedics \\
S6  & Male   & 26 & Resident & Student & Routine & Orthopedics \\
S7  & Female & 27 & Resident & Student & Routine & Orthopedics \\
S8  & Male   & 32 & Resident & Student & On-demand   & Orthopedics \\
S9  & Female & 28 & Resident & Student & Routine & Orthopedics \\
S10 & Male   & 29 & Resident & Student & On-demand   & Orthopedics \\
S11 & Male   & 31 & Resident & Student & Routine & Orthopedics \\
S12 & Female & 25 & Intern & Student & On-demand   & Orthopedics \\
S13 & Male   & 27 & Resident & Student & Routine & Orthopedics \\
S14 & Female & 26 & Resident & Student & Routine     & Orthopedics \\
S15 & Male   & 28 & Resident & Student & On-demand   & Orthopedics \\
S16 & Female & 30 & Resident & Student & Routine     & Orthopedics \\
S17 & Male   & 30 & Resident & Student & On-demand   & Orthopedics \\
S18 & Female & 29 & Resident & Student & Routine     & Orthopedics \\
S19 & Female & 23 & Intern & Student & Routine     & Orthopedics \\
\hline
\end{tabular}
\end{table}

\begin{table}[h!]
\begin{center}
\caption{Demographic information of senior physicians in the controlled study: PBL participation encompasses various levels of engagement, including regular involvement (Routine) and participation based on individual needs (On-demand).}
\label{tab:between_teachers}
\begin{tabular}{p{0.1cm}p{0.8cm}p{0.3cm}p{1.3cm}p{0.8cm}p{1.6cm}p{1.4cm}}
\hline
\textbf{ID} & \textbf{Gender} & \textbf{Age} & \textbf{Title} & \textbf{PBL Role} & \textbf{PBL Participation} & \textbf{Clinical Domain} \\
\hline
T1 & Male   & 43 & Associate Chief Physician & Teacher & Routine     & Orthopedics \\
T2 & Male   & 41 & Attending & Teacher & On-demand   & Orthopedics \\
\hline
\end{tabular}
\end{center}
\end{table}

\subsubsection{Baseline}
\par \revision{The baseline system was designed based on several key considerations. First, its application scenario needs to align with individual analysis in PBL, as opposed to traditional PBL teaching scenarios or retrospective learning that involves comprehensive knowledge review. Second, the analysis process should follow a structured PBL approach, rather than a ``strawman'' format that presents materials without organization.}

\par \revision{Based on these criteria, our baseline system was developed by removing analysis features from \textit{e-MedLearn}, such as the diagnosis list, QA section, and mind map. The baseline retains general fixed-format prompts like ``\textit{Patient Information, Inferences/Hypotheses, Proposed Actions, and Issues for Further Learning}'' to guide the analysis process. In \textbf{Stage 2: Information Analysis} and \textbf{Stage 3: Record and Review}, the baseline system provides fixed-format questions to guide users' analysis through the process and reflection. An example interface of the baseline system is shown in \autoref{fig:baseline}.}


\begin{figure*}[h]
    \centering
\includegraphics[width=\textwidth]{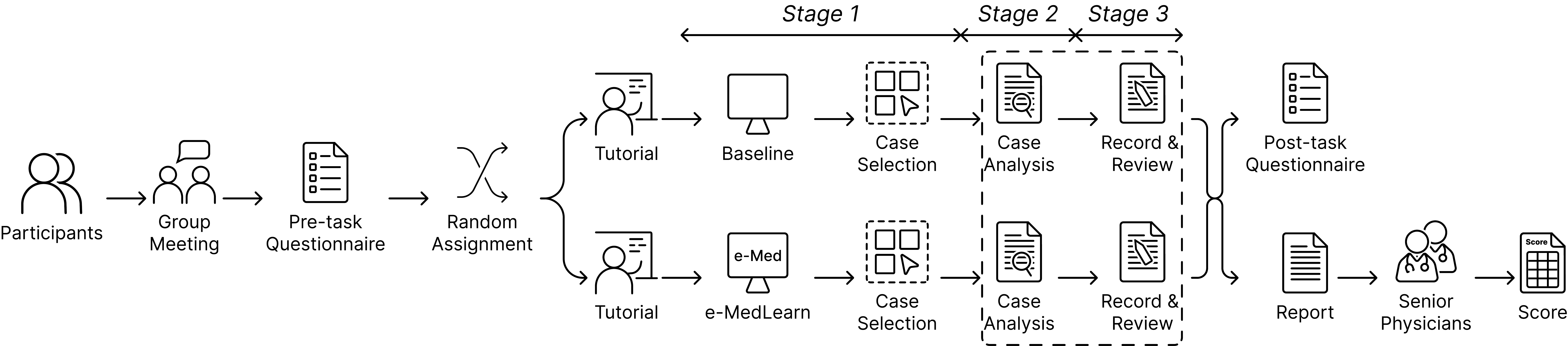}
\caption{\revision{The pipeline of the controlled study. Participants were randomly assigned to two conditions: the baseline and \textit{e-MedLearn}. In \textbf{Stage 1: Data Collection}, participants browsed and selected orthopedic cases of interest using the system. In \textbf{Stage 2: Information Analysis}, all participants analyzed the same case. In \textbf{Stage 3: Record and Review}, participants documented and reviewed their analyses. Following these stages, participants completed a post-task questionnaire and submitted a case analysis report, which was then evaluated by senior physicians.}}
\label{fig:controlled_study}
\end{figure*}

\subsubsection{Procedure}
\par \revision{Prior to the study, we conducted a 15-minute group meeting to ensure all participants shared a clear understanding of PBL learning and clinical reasoning-based differential diagnosis (as described in \autoref{fig:workflow}). We also outlined the experiment's tasks, which involved selecting cases through the PBL process, followed by case analysis and review. Participants first completed a pre-questionnaire to collect demographic information, including their specialty and prior PBL participation. They were then randomly assigned to one of two groups: $10$ in the \textit{e-MedLearn} condition and $9$ in the baseline condition. Afterward, we provided system tutorials tailored to each group, along with sample case data, to help participants familiarize themselves with the system.}

\par \revision{The study is based on the three-stage pipeline of \textit{e-MedLearn}. In \textbf{Stage 1: Data Collection}, users begin by selecting orthopedic cases of interest for about 20 minutes. This step primarily evaluates the system's ability to identify and select relevant cases. To evaluate the system's application of clinical reasoning in specific case-based PBL analysis through comparison, we provided all users with the same case at the beginning of \textbf{Stage 2: Information Analysis}. Participants used the system for approximately 50 minutes to complete the case analysis and subsequent record and review process in \textbf{Stage 3: Record and Review}.}

\par \revision{After completing the analysis, participants filled out a post-task questionnaire and conducted a 10-minute interview to share their views of the questions in the questionnaire. They also completed a templated case analysis report (\autoref{tab:report_template}), documenting their analysis and reasoning process following the Clinical Reasoning Cycle (CRC) framework~\cite{james2021levette}. In the \textit{e-MedLearn} condition, participants could directly review the analyzed information within the system while preparing their case analysis report. In contrast, participants in the baseline condition relied on fixed-format prompting questions to review their notes. The reports were then evaluated by two senior physicians (T1, T2), who scored them blindly based on the completeness and organization of the clinical reasoning process. Detailed scoring criteria are provided in \autoref{tab:report_criteria}. Prior to scoring, a meeting was held to ensure the senior physicians have a consistent understanding of the criteria. The final score for each report was calculated as the average of the two physician's scores. This study was approved by the institutional IRB, and each participant received a \$20 compensation for their time and effort.}

\subsubsection{Data Collection and Measurement}
\par \revision{We collected the questionnaires and case analysis reports from the participants and conducted a quality check on the data. One participant's data in the \textit{e-MedLearn} condition was excluded due to an analysis duration of less than $30$ minutes. In the end, we obtained $18$ valid data sets, with $9$ from the \textit{e-MedLearn} condition and $9$ from the baseline condition.}

\par \revision{For the post-task questionnaire, we used a 7-point Likert scale (1: Strongly disagree, 7: Strongly agree) to collect participants' feedback on the analysis process and their perceptions of their own performance. First, under \textbf{Usability}, we adapted questions from the System Usability Scale (SUS)~\cite{brooke1996sus} to assess (1) \textit{Ease of use}, (2) \textit{Helpfulness for clinical reasoning}, (3) \textit{Satisfaction}, (4) \textit{Distraction}, and (5) \textit{Recommendation and future use}. Second, to assess \textbf{Workload}, we based questions on the NASA-TLX survey~\cite{hart1988development}, focusing on (1) \textit{Mental demand}, (2) \textit{Physical demand}, (3) \textit{Temporal demand}, (4) \textit{Performance perception}, (5) \textit{Effort}, and (6) \textit{Frustration}. Third, we designed questions (\autoref{tab:effectiveness}) to evaluate the \textbf{Effectiveness} of the system, including (1) \textit{Targeted case selection}, (2) \textit{Thoughtful information identification}, (3) \textit{Structured organization}, (4) \textit{Evidence-based reasoning}, (5) \textit{Reasoning-supported action plan}, (6) \textit{Prognosis consideration}, (7) \textit{Clinical reasoning reflection}. Due to the small sample size, we applied the nonparametric Mann-Whitney U test~\cite{mann1947test} to compare results between conditions, as it does not assume specific data distribution. For the case analysis reports, we performed Spearman's Rank Correlation~\cite{spearman1904proof} to assess consistency between the two raters' scores. It is important to note that the case analysis report primarily evaluates the user's ability to organize and reflect on the case analysis process based on clinical reasoning in the PBL context, rather than immediate knowledge acquisition.}

\begin{table*}
  \centering
  \caption{\revision{Questions about the Effectiveness in the post-task questionnaire.}}
  \label{tab:effectiveness}
  \begin{tabular}{lp{0.6\textwidth}}
\hline
(1) Targeted case selection & I can select cases suitable for in-depth analysis based on my learning objectives. \\
\hline
(2) Thoughtful information identification & I can identify key diagnostic information through thoughtful exploration. \\
\hline
(3) Structured organization & I can effectively organize and integrate the collected information for analysis. \\
\hline
(4) Evidence-based reasoning & My reasoning process is clear and logical based on evidence. \\
\hline
(5) Reasoning-supported action plan & My action plans are supported by reasoning with a clear rationale. \\
\hline
(6) Prognosis consideration & I can recognize the prognosis and consider relevant follow-up actions. \\
\hline
(7) Clinical reasoning reflection & I can reflect on my reasoning process and identify insights into its strengths and weaknesses. \\
\hline
  \end{tabular}
\end{table*}

\subsection{Testing Interviews}
\par \revision{We subsequently conducted testing interviews to collect feedback on participants' practice. To avoid any influence from the prior study, we re-recruited 10 novice physicians (7 males, 3 females, average age: 26.6) through snowball mailing lists. We also involved 3 senior physicians (2 males, 1 female, average age: 46.7), where two physicians scored the case analysis reports in the controlled study but had no direct exposure to the system. All of the participants were not co-authors of this work and did not participate in \textbf{Phase One}. Participant details are provided in \autoref{tab:participants}.}


\subsubsection{Procedure}
\par The process of the testing interviews was structured as follows: first, reviewing current PBL practices and participants' perceptions; second, introducing the system's design considerations and features aimed at enhancing diagnostic reasoning during the PBL process; and third, collecting feedback on the \revision{fully implemented system} after its use in the PBL process. This study was approved by the institutional IRB, and each participant received \$20 in compensation for their time and participation.

\par We began by exploring participants' current practices and perceptions of PBL in educational settings, with a focus on the organization and knowledge acquisition processes. \revision{In PBL activities, novice physicians typically engage as students, using the sessions for learning, while senior physicians serve as instructors, guiding and facilitating the process.} For novice physicians, we focused on their experience as students in the PBL process, as they represent the primary target users of \textit{e-MedLearn}. For senior physicians, we examined their roles as instructors, particularly their perspectives on PBL teaching and their observation of students' progress throughout the process.

\par Next, we introduce the system's design considerations, highlighting how it supports users in applying knowledge to real-world scenarios and enhances their PBL skills, particularly in clinical reasoning and differential diagnosis. At this stage, we do not present the system itself.

\par Following the introduction, the system is presented to the users, and its features are explained. Users are then given $5$ minutes to familiarize themselves with the system before proceeding with a full PBL analysis, which includes case selection, analysis, and reflection, taking approximately $50$ minutes. Upon completing the analysis, interviews are conducted to collect feedback on the system.

\par Due to the senior physicians' experience and roles, they are not required to fully engage in the entire PBL analysis process. Instead, they focus on understanding the system's high-level design considerations and provide feedback based on their interactions and perspectives during the subsequent interviews.

\begin{table}[h!]
\caption{Demographic information of participants in the testing interviews: PBL participation encompasses various levels of engagement, including regular involvement (Routine), participation based on individual needs (On-demand), and temporary involvement for specific purposes (Ad-hoc).}
\label{tab:participants}
\begin{tabular}{p{0.2cm}p{0.8cm}p{0.3cm}p{1.3cm}p{0.8cm}p{1.6cm}p{1.5cm}}
\hline
\textbf{ID} & \textbf{Gender} & \textbf{Age} & \textbf{Title} & \textbf{PBL Role} & \textbf{PBL Participation} & \textbf{Clinical Domain} \\
\hline
P1  & Male   & 26 & Resident & Student & Routine & Orthopedics \\
P2  & Female & 25 & Intern   & Student & On-demand   & Neurosurgery \\
P3  & Male   & 28 & Resident & Student & Routine & Neurosurgery \\
P4  & Female & 25 & Intern   & Student & On-demand   & Neurosurgery \\
P5  & Male   & 27 & Resident & Student & Routine & Neurosurgery \\
P6  & Male   & 26 & Resident & Student & Routine & Orthopedics \\
P7  & Female & 28 & Resident & Student & On-demand & Neurosurgery \\
P8  & Male   & 26 & Resident & Student & Routine & Orthopedics \\
P9  & Female & 28 & Resident & Student & Routine & Neurosurgery \\
P10 & Male   & 27 & Resident & Student & On-demand   & Neurosurgery \\
P11 & Male   & 43 & Associate Chief Physician & Teacher & Routine & Orthopedics \\
P12 & Female & 38 & Attending & Teacher & Ad-hoc & Neurosurgery \\
P13 & Male   & 41 & Attending & Teacher & On-demand   & Orthopedics \\
\hline
\end{tabular}
\end{table}

\subsubsection{Data Collection and Analysis}
\par Each interview session was conducted via video conference software. With the participants' knowledge and consent, we audio and video recorded the entire testing interview process. The qualitative data from these recordings were analyzed using thematic analysis~\cite{braun2006using,braun2012thematic} and affinity diagram\cite{corbin2014basics}.

\par \revision{First, the audio recordings were transcribed into text using Otter.ai~\cite{otterai_website}, an automated speech transcription tool.} Subsequently, the first author manually verified the transcripts to ensure accuracy. The transcripts were then organized preliminarily. Two team members conducted an inductive analysis to identify various themes, sub-themes, and relevant codes.

\par The analysis includes reflections from both students and instructors on the current PBL approach, as well as their thoughts and feedback on the system's design and features. These insights highlight opportunities for \textit{e-MedLearn} to enhance clinical reasoning-based differential diagnosis within the PBL process. Additionally, in-task questionnaires were collected to evaluate the feature effectiveness from participants with the role of students role in the PBL process, as the \textbf{PBL Role} shown in \autoref{tab:participants}. The questions in the in-task questionnaires are shown in the Appendix.

\section{Results}
\subsection{Results of Controlled Study}
\par We present the results of the controlled study, including feedback from the post-task questionnaire and scores from the case analysis report. \textit{Mean} and \textit{SD} represent the average value and standard deviation for each question within the respective groups. \textit{M} indicates the Mann-Whitney U value, and \textit{p} denotes the level of statistical significance.

\begin{figure}[h]
\centering
\includegraphics[width=\columnwidth]{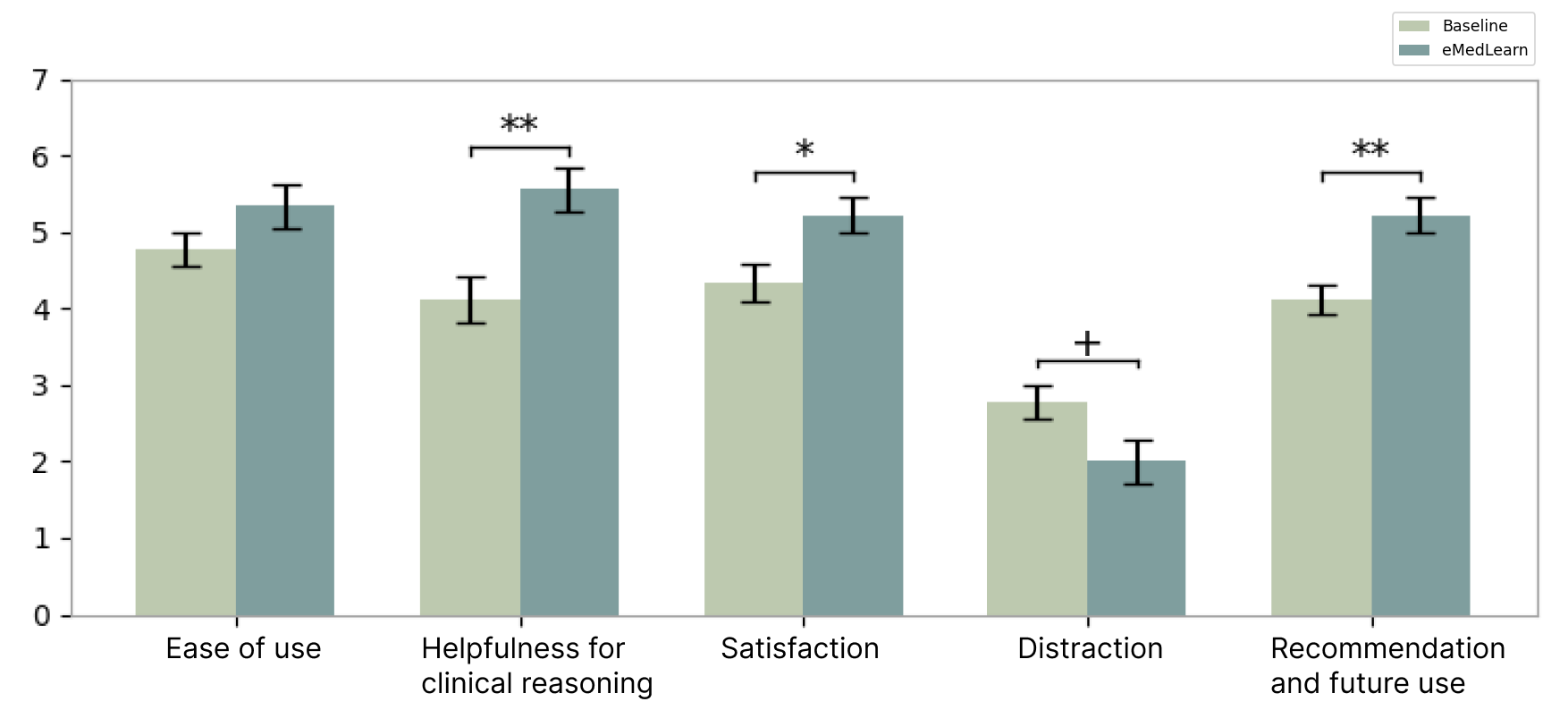} 
\caption{\revision{The results of usability perceived by users (*: \( p < 0.05 \); **: \( p < 0.01 \); ***: \( p < 0.001 \)).}}
\label{fig:Usability}
\end{figure}

\par \revision{\textbf{Usability:} The usability results are shown in \autoref{fig:Usability}. Compared to the baseline system, participants generally provided more positive feedback regarding the usability of \textit{e-MedLearn}. Specifically, users rated the \textit{Ease of use} of \textit{e-MedLearn} (\textit{Mean} = 5.33, \textit{SD} = 0.86) higher compared to the baseline (\textit{Mean} = 4.78, \textit{SD} = 0.67). ``\textit{The design of steps in the e-MedLearn system is reasonable and aligns with our diagnostic analysis process, so it didn't add any complexity that would make it feel hard to use.} (S3)'' Furthermore, participants in \textit{e-MedLearn} condition scored significantly higher on \textit{Helpfulness for Clinical Reasoning} (\textit{M} = 11, \textit{p} < 0.01) and \textit{Recommendation for Future Use} (\textit{M} = 10, \textit{p} < 0.01). For the scores of \textit{Distraction}, \textit{e-MedLearn} (\textit{Mean} = 2, \textit{SD} = 0.87) also received lower scores compared to the baseline (\textit{Mean} = 2.78, \textit{SD} = 0.67), while this difference was not statistically significant. Additionally, \textit{e-MedLearn} demonstrated significantly higher scores for \textit{Satisfaction} (\textit{Mean} = 5.22, \textit{SD} = 0.67) compared to the baseline system (\textit{Mean} = 4.33, \textit{SD} = 0.71) (\textit{M} = 16, \textit{p} < 0.05), indicating a higher level of overall usability satisfaction. Users acknowledged that the system for helping``\textit{make their analysis more concrete, with each reasoning step based on the current evidence and grounded in a clear rationale.} (S7)''}

\par
\begin{figure}[h]
\centering
\includegraphics[width=\columnwidth]{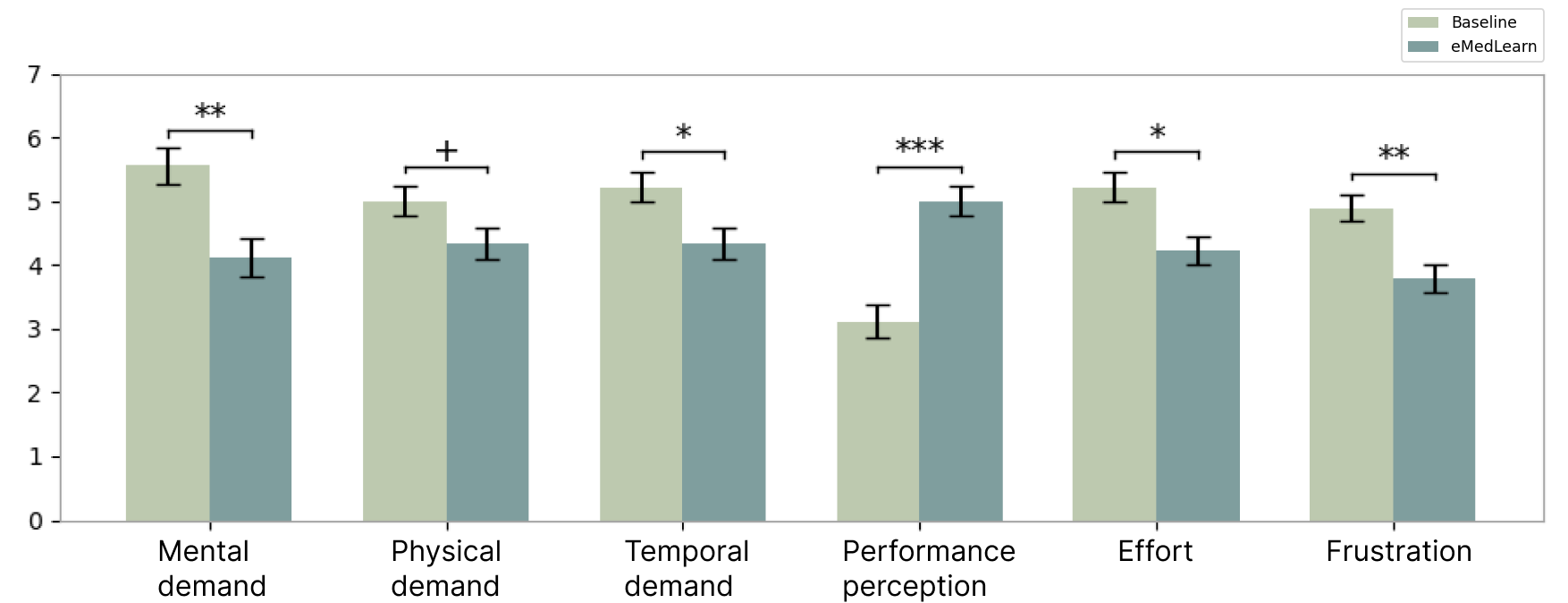} 
\caption{\revision{The results of workload perceived by users (*: \( p < 0.05 \); **: \( p < 0.01 \); ***: \( p < 0.001 \)).}}
\label{fig:Workload}
\end{figure}

 \revision{\textbf{Workload:} The workload results are shown in \autoref{fig:Workload}. Overall, participants in the \textit{e-MedLearn} condition reported significantly lower workload perceptions across several dimensions, including \textit{Mental Demand} (\textit{M} = 70, \textit{p} < 0.01), \textit{Temporal Demand} (\textit{M} = 65, \textit{p} < 0.05), and \textit{Effort} (M = 68, p < 0.05). Participant noted that ``\textit{I provided a low mental demand score because I think the system made it easier for me to form evidence-based hypotheses with corresponding rationales, thereby supporting reasoning process.} (S8)'' Participants in \textit{e-MedLearn} condition also reported lower \textit{Physical Demand} (baseline: \textit{Mean} = 5.56, \textit{SD} = 0.88; \textit{e-MedLearn}: \textit{Mean} = 4.11, \textit{SD} = 0.93), with no statistically significant difference. Additionally, \textit{e-MedLearn} participants reported significantly lower \textit{Frustration} scores (\textit{Mean} = 3.78, \textit{SD} = 0.67) compared to the baseline condition (\textit{Mean} = 4.89, \textit{SD} = 0.60) (\textit{M} = 71, \textit{p} < 0.01). ``\textit{The fixed questions in the <baseline> system didn’t really help with my exploratory thinking, so I felt unsure and not very satisfied with the analysis process.} (S4)''}

\par
\begin{figure*}[h]
\centering
\includegraphics[width=\textwidth]{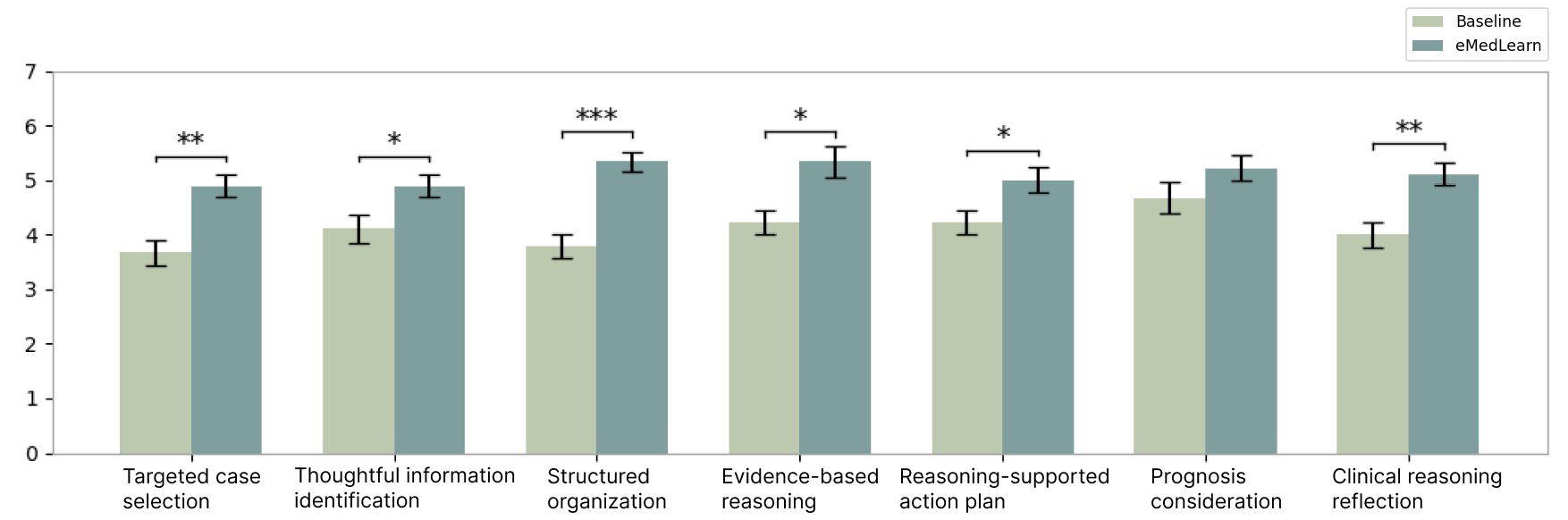}
\caption{\revision{The results of effectiveness perceived by users (*: \( p < 0.05 \); **: \( p < 0.01 \); ***: \( p < 0.001 \)).}}
\label{fig:Effectiveness}
\end{figure*}

 \revision{\textbf{Effectiveness:} \autoref{fig:Effectiveness} presents the results on effectiveness. Overall, the \textit{e-MedLearn} condition achieved significantly higher scores. In \textbf{Stage 1: Data Collection}, \textit{e-MedLearn} (\textit{Mean} = 4.89, \textit{SD} = 0.60) outperformed the baseline (\textit{Mean} = 3.67, \textit{SD} = 0.71) in \textit{Targeted case selection} with a significant difference. In \textbf{Stage 2: Information Analysis}, \textit{e-MedLearn} demonstrated significantly better performance in \textit{Thoughtful information identification} (\textit{M} = 19, \textit{p} < 0.05), \textit{Structured organization} (\textit{M} = 3, \textit{p} < 0.001), \textit{Evidence-based reasoning} (\textit{M} = 13, \textit{p} < 0.05), and \textit{Reasoning-based action plans} (\textit{M} = 18.5, \textit{p} < 0.05). ``\textit{Problem-based analysis helps me think critically, like understanding why I make certain decisions and what evidence backs them up. I believe a good diagnosis really needs this kind of thinking to figure out both the disease and the action plan.} (S9)'' In the \textbf{Stage 3: Record and Review}, \textit{e-MedLearn} (\textit{Mean} = 5.22, \textit{SD} = 0.67) showed improved facilitation of \textit{Prognosis consideration} compared to the baseline (\textit{Mean} = 4.67, \textit{SD} = 0.87), with feedback indicated that ``\textit{the prompting questions helped me think through relevant issues, making sure I didn’t miss any important prognostic details and keeping the diagnostic process thorough.}''. Additionally, \textit{e-MedLearn} (\textit{Mean} = 5.11, \textit{SD} = 0.60) significantly outperformed the baseline (\textit{Mean} = 4.0, \textit{SD} = 0.70) in promoting \textit{Clinical reasoning reflection} (\textit{M} = 10.5, \textit{p} < 0.01). As the participants indicated, ``\textit{After the analysis, I mainly reflect on the reasoning process to see where things went wrong. The diagnosis list and mind map help me figure this out easily.} (S3)''}

\par \revision{We compared the case analysis reports submitted by users across eight dimensions, using the criteria in \autoref{tab:report_criteria}. For each dimension, we performed a Mann-Whitney U test to compare the scores between the baseline and \textit{e-MedLearn} conditions. As shown in \autoref{tab:mann_whitney_results}, the report scores revealed significant differences in different dimensions of clinical reasoning. These results highlight the effectiveness of \textit{e-MedLearn} in facilitating clinical reasoning during PBL case analysis.}

\begin{table}[h!]
\centering
\caption{\revision{The average scores between Baseline and \textit{e-MedLearn} condition in different dimensions of case analysis report criteria. "U" indicates Mann-Whitney U test results, and "p" indicates p-values for statistical significance (*: \( p < 0.05 \); **: \( p < 0.01 \); ***: \( p < 0.001 \)).}}
\label{tab:mann_whitney_results}
\resizebox{\columnwidth}{!}{
\begin{tabular}{lcccc}
\hline
 & \textbf{Baseline} & \textbf{e-MedLearn} & \textbf{U} & \textbf{p} \\
\hline
Consider the Patient Situation & 3.11 & 5.33 & 2.00 & *** \\
\hline
Collect Cues and Information & 4.00 & 5.67 & 7.50 & ** \\
\hline
Process Information & 3.22 & 4.78 & 7.50 & ** \\
\hline
Identify Problems/Issues & 3.22 & 4.44 & 17.00 & * \\
\hline
Establish Goals & 3.78 & 5.78 & 8.50 & ** \\
\hline
Take Action & 3.78 & 5.00 & 15.00 & * \\
\hline
Evaluate Outcomes & 3.67 & 5.22 & 9.50 & ** \\
\hline
Reflect on the Process and New Learning & 3.00 & 5.22 & 4.00 & ** \\
\hline
\end{tabular}
}
\end{table}

\subsection{Results of Testing Interviews}
\label{testing_interviews}
\par
\begin{figure}[h]
\centering
\includegraphics[width=\columnwidth]{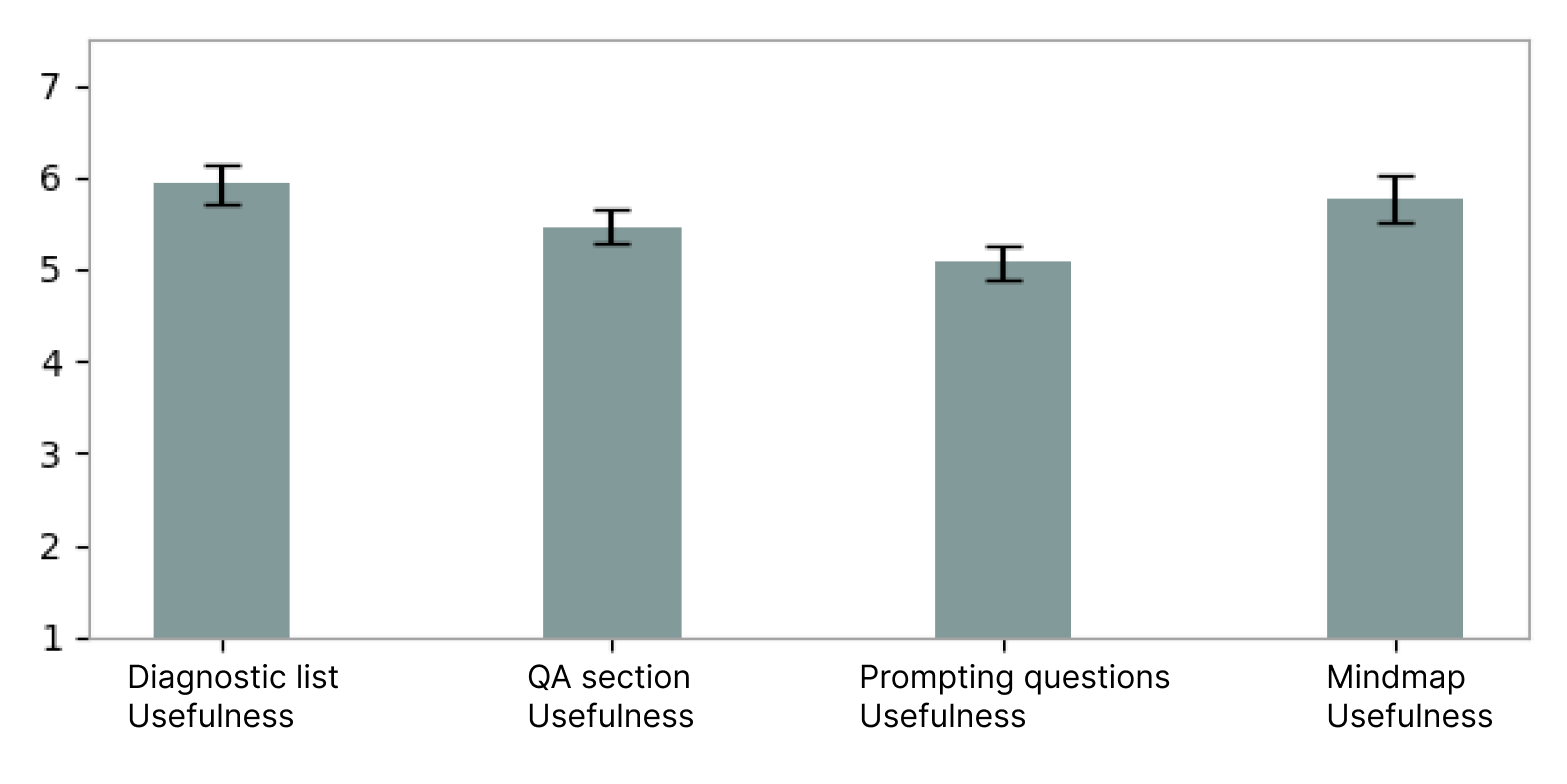}
\caption{\revision{The usefulness of different features in \textit{e-MedLearn}.}}
\label{fig:usefulness}
\end{figure}

 For the in-task questionnaire \revision{in the \textbf{testing interviews}}, we used a seven-point Likert scale (1: Strongly disagree, 7: Strongly agree) to assess the perceived usefulness of system features. As shown in \autoref{fig:usefulness}, most features in \textit{e-MedLearn} were recognized as useful, though the prompting questions received relatively lower scores (\textit{Mean} = 5.08, \textit{SD} = 0.64).

\par For the qualitative data \revision{in the \textbf{testing interviews}}, we employed a thematic analysis approach~\cite{braun2006using,braun2012thematic}, categorizing the findings into three key themes: motivation and considerations in transitioning from knowledge to practice, the application of evidence-based reasoning in analysis, and the benefits observed in development and improvement. A detailed discussion of each theme is presented below.

\noindent \textbf{Theme 1: ``\textit{So gaps exist between problem-solving and patient diagnosis.}'' The motivation and considerations from knowledge to practice}

\subsubsection{Awaken self-directed identification and awareness.} Regarding the role of \textit{e-MedLearn} in bridging knowledge and practice, participants recognized that the system offers a solution for selecting PBL cases tailored to their learning status. They highlighted that this approach awakens their awareness during self-directed learning, including the identification of both knowledge mastery and learning intentions as they transition from theory to practice through PBL.
\begin{quote}
``\textit{Sometimes we try to simulate PBL for practice based on our own needs, but we find it challenging to identify the right cases without proper guidance. I think one of the key things about the system is that it helps us identify our learning progress and gives us the support to make informed case choices based on our specific interests and focus areas.}'' (P5)
\end{quote}


\par The feedback also reflected the system's effectiveness in clarifying learning intentions during the case selection process. Participants noted that previous challenges in selecting cases during PBL often stemmed from an unclear understanding of learning objectives. Features such as case display by incidence rate, logical chain, and relevance to classic diseases have been instrumental in enhancing users' awareness, thereby providing a solid foundation for a self-directed PBL process.


\begin{quote}
``\textit{When I saw this <difficulty reference for each case>, it struck me as exactly what I need for real clinical practice. In the early stages of learning, we concentrate on classic disease cases to build a solid foundation. As we advance, we shift to more complex cases with longer logical chains to sharpen our diagnostic reasoning skills with that knowledge.}'' (P9).
\end{quote}

\noindent \textbf{Theme 2: ``\textit{\revision{My analysis process fails due to unsubstantiated reasoning.}}'' The application of evidence-based reasoning in analysis}
\subsubsection{The acquisition and extension of evidence.} Participants emphasized that the key to applying learned knowledge in real-world diagnosis lies in reasoning ability. Rather than merely analyzing all available evidence, this reasoning process involves making initial judgments based on presented evidence and then uncovering additional information. This approach mirrors real clinical scenarios, where, as one participant noted, ``\textit{patients typically start by describing their symptoms, rather than providing all relevant information upfront} (P11).'' It is through the interaction between physician and patient that further information is gathered, which is then used for exclusion and differential diagnosis. The ability to effectively gather this information is a crucial aspect of diagnostic training. Participants acknowledged that the explicit logical reasoning features of \textit{e-MedLearn} assist them significantly in this process.

\begin{quote}
\textit{``I understand that in real clinical practice, unlike solving problems where you have all the information upfront, physicians gather more information through conversation and questioning with the patient, and make judgments based on this information. The system, by providing a question-and-answer feature that reveals information gradually as inquiries are made, rather than displaying all information at once, simulates this process, making my thinking more aligned with clinical diagnostic reasoning}. (P7)''
\end{quote}

\par The physicians further elaborated that the practice of expanding and analyzing based on initial evidence is integral to developing students' reasoning and analytical skills. Senior physicians, while conducting patient interviews, simultaneously build logical connections in their minds—a skill refined through experience and repeated practice. Notably, when they document `negative symptoms'---those that the patient did not exhibit in the medical record, it demonstrates what one participant referred to as a ``medical way of thinking'' (P4).

\begin{quote}
``\textit{These denied symptoms mean that the patient was asked about them, but they said no, rather than the doctors overlooking them. The system simulates this process much like how a teacher guides responses in PBL teaching, helping students develop their medical thinking and reasoning skills.} (P13)''
\end{quote}%

\par Additionally, the senior physician pointed out that \revision{students} may tend to swing to the opposite extreme, asking about numerous symptoms one by one without considering the reasoning behind their questions. A clinical reasoning exemplar should strike a balance by ``\textit{avoiding redundancy while ensuring all essential aspects are covered}'' in the inquiry process (P12). Moving forward, we aim to analyze users' questioning patterns to guide them toward more systematic and logical approaches to patient evaluation.

\subsubsection{Evidence prompts clinical reasoning and differential diagnosis.} Almost all participants emphasized the importance of evidence-based reasoning in diagnosis. They noted that the diagnostic process entails forming initial hypotheses based on available evidence and subsequently updating these hypotheses as new information emerges, ultimately leading to a final diagnosis. Feedback indicates that these considerations are highly valued in enhancing the PBL process.

\begin{quote}
``\textit{In PBL activities, I often found that \revision{my analysis process failed due to unsubstantiated reasoning,} as I relied too much on intuition and lacked systematic thinking. With this system, the structured analytical framework it provides is both clear and practical, helping me become more systematic and logical in my approach.} (P9)''
\end{quote}%

\par As noted by several participants, systematically engaging users in explicit, evidence-based reasoning enhances their medical literacy on both cognitive and methodological levels. One participant remarked, ``\textit{When you consistently record and organize evidence and thoughts in a structured way, such as following a framework like <FILA>, your analysis process becomes more orderly and coherent}'' (P8). Additionally, participants observed that this evidence-based approach aids in adopting the patient's perspective, thereby improving the efficiency of information gathering during patient interactions and enhancing the clarity of explanations throughout the communication process.

\begin{quote}%
``\textit{Patients sometimes provide irrelevant information during conversations with physicians. The evidence-based understanding during diagnosis helps us guide the patient to focus on their condition, facilitating our diagnostic reasoning. It also helps us reasonably explain our requests and decisions when the patient is confused or in doubt.} (P5)''
\end{quote}%

\noindent \textbf{Theme 3: ``\textit{What reflections should I make after practice?}'' The benefits gained in development and improvement}
\subsubsection{Post-diagnosis review and reflection.} In the interviews, participants noted that \textit{e-MedLearn} supports learning about prognosis following diagnosis, an essential aspect of a physician's responsibilities. They highlighted that the system's features enhance the comprehensiveness of diagnostic analysis and help address the ``\textit{potential neglect of prognosis considerations due to an excessive focus on correctness and reasoning once the diagnosis is revealed}'' (P11).

\begin{quote}
``\textit{After the analysis of case data <in PBL process>, I spontaneously became curious about how my analysis matched the actual results and focused on the outcome. Looking back, I realize that I often neglected the subsequent <post-diagnosis> content. The prompting questions provided by the system help remind me to continue paying attention to the case after the diagnosis and to record key points as part of my ongoing learning.}'' (P2)
\end{quote}%

\par Regarding the benefits derived from the analysis, both \revision{students and teachers} underscored the importance of review and reflection following the learning process, highlighting that these aspects are as crucial as the analysis itself. Feedback revealed that less experienced students may experience confusion after completing the analysis and struggle with the question of, ``\textit{What reflections should I make after practice?}'' (P6).

\begin{quote}
``\textit{After finishing the diagnostic analysis, the answers usually get revealed, but the feedback from teachers is often just a general summary of the whole process. It rarely zeroes in on individual analysis and errors. So, I used to struggle with finding ways to improve through reflection, especially when I wasn’t very skilled at the analysis process.}'' (P2)
\end{quote}%

\par Participants observed that \textit{e-MedLearn} effectively bridges the gap between practice and reflection in the PBL process. They noted that the system offers a structured workflow that guides users from analysis to documentation, allowing them to review their reasoning based on the thought process rather than just the outcome. This approach facilitates high-level, structured reflection and fosters continuous improvement beyond the specifics of individual cases.

\begin{quote}
``\textit{In e-MedLearn, we can see the previous diagnostic list that records all hypotheses and corresponding thoughts, as well as the mind map that organizes and records information. This allows me to review the entire traceable analysis from the perspective of my thought process. Correcting thinking patterns rather than merely focusing on the outcome is undoubtedly more valuable.}'' (P1)
\end{quote}

\par From a long-term development perspective, the interviews indicate reflection on methodology and thought processes is more beneficial for enhancing learners' abilities. With the previous perception of improvement in current PBL activities, participants consider the structured organization through \textit{e-MedLearn} could better facilitate valuable reflections for improvement and also is ``\textit{consistent with the intention of PBL activities that transition knowledge into clinical practice}'' (P9).

\section{DISCUSSION}
\par To promote the application of theoretical knowledge in practical settings, we explored the integration of clinical reasoning and differential diagnosis methodologies into PBL. Our work focused on advancing medical students' skills throughout the PBL process, with a strong emphasis on evidence-based discovery, exploration, and reasoning. Based on these findings, we identified design implications for enhancing the PBL experience, particularly in the application of diagnostic methodologies, while considering both educational and clinical scenarios. Additionally, we explored the potential for fostering continuous clinical development and comprehensive diagnostic decision-making within medical education.

\subsection{Implication for Design in PBL Practice}
\subsubsection{Focusing on Clinical Diagnostic Methodologies}
\par Our study's findings suggest that \textit{e-MedLearn} effectively bridges the gap between current PBL practices and methodology-driven clinical reasoning and differential diagnosis. These insights provide valuable guidance for HCI and AI-based design considerations in the PBL process, facilitating the transition from theoretical knowledge to practical application.

\par \textbf{Promoting evidence-based divergent thinking rather than uncritical usage.} Participants acknowledged that \textit{e-MedLearn} encourages users to actively explore and gather evidence by avoiding the presentation of all case data at once. This approach fosters the development of students' medical reasoning through directed exploration. However, some users reported experiencing a ``cold start'' during the data exploration process, where the expected divergent thinking based on initial evidence did not unfold as intended.
\par A key concern raised by participants is the system's step-by-step inquiry of evidence exploration limits in providing sufficient methodological guidance for informed exploration. While general prompts such as ``\textit{What symptoms should be inquired about next?}'' may encourage convergent analysis~\cite{voss2022comparing}, they are sometimes viewed as inadequate for promoting creative divergent thinking~\cite{Razumnikova2012,baer2014creativity}. A potential solution is to introduce prompts more closely tied to the data, such as ``\textit{Based on the symptom, you may need to inquire about <related aspects>?}'' However, this should be approached carefully to avoid explicit instructions that could reveal the content, which may increase learners' cognitive load in an information-rich context~\cite{de2010cognitive} and hinder the directed analysis.


\par \textbf{Emphasizing problem-driven reasoning.} Interview feedback emphasized the critical role of problem-based reasoning during the analysis process. Senior physicians (P11, P13) noted that ``\textit{the essence of case analysis in PBL is the continuous process of raising questions and solving problems based on those questions}'', which aligns with the process of reasoning and differential diagnosis in clinical practice~\cite{corazza2021diagnostic}. Although framework-based prompting questions have proven effective in advancing problem-motivated analysis through diagnostic methodologies, some participants expressed a desire for more flexible and targeted prompting questions to further enhance the analysis process. Unlike lecture-based learning~\cite{antepohl1999problem}, which focuses on content delivery, and case-based learning~\cite{williams2005case}, which involves retrospective case reviews, PBL is characterized by its emphasis on problem-driven reasoning that bridges knowledge and practice. This necessitates careful attention to system design, not only to support problem analysis but also to foster students' skills in recognizing and formulating problems, a key aspect echoed in tools designed for academic analysis~\cite{yuan2023critrainer,peng2022crebot,sun2024reviewflow}. In this context, AI's role extends beyond facilitating evidence exploration to guiding reasoning. A potential solution is leveraging large language models (LLMs) to suggest possible areas of exploration based on user input, thereby enhancing the problem-driven reasoning process.

\subsubsection{Bridging Medical Education with Clinical Practice}
\par The design considerations of \textit{e-MedLearn} demonstrate a theoretical approach to promoting the application of PBL in clinical diagnostics through clinical reasoning. However, in addition to this theoretical methodology, it is essential to account for various real-world factors present in actual clinical environments. Feedback obtained from the study highlights the importance of integrating educational considerations with practical diagnostic processes.

\par \textbf{Integrating design consideration with clinical workflow.} While participants recognized the system's support in facilitating the PBL analysis process, they also highlighted the gap between theoretical design considerations and practical clinical scenarios. A common observation was the need to adapt to real-world clinical environments, where factors such as patient interaction, emotional stress, and distractions complicate communication and shared decision-making (SDM)~\cite{ley1988communicating,stub2017complementary,hargraves2016shared,hao2024advancing}. Physicians emphasized that in clinical practice, these factors play a critical role. While classroom-based PBL methods offer foundational learning, formats such as Virtual Patients (VP) and Standardized Patients (SP)~\cite{flanagan2023standardized,malik2024virtual} are crucial for preparing students to navigate real-world complexities. However, placing too much emphasis on practicality could limit the understanding of underlying methodologies. Therefore, integrating design considerations with clinical workflows involves striking a balance between practical experience and the mastery of theoretical approaches. Future efforts should focus on how to seamlessly embed methodological training into clinical environments. Advanced technologies like AI and LLM, combined with VR/MR simulations, could play a significant role in enhancing immersive learning and practice, offering varied patient role simulations and enriching real-world preparation.

\par \textbf{Leveraging evidence as the explanation in diagnostic reasoning.} Participants in the study highlighted the importance of diagnostic updates informed by evidence. From a logical reasoning standpoint, this reflects the differential diagnosis process in evidence-based medicine~\cite{akobeng2005principles}, where reasoning and analysis are rooted in evidence. In this context, evidence can be seen as a form of explanation, helping physicians construct reasoning that supports their cognitive decisions. This approach, focused on learning to leverage self-reinforcing reasoning as a form of explanation, aligns with clinical practice as metacognition~\cite{cloude2022role}, which develops through experience and enhances the efficiency of outpatient processes. In PBL, utilizing the explanatory nature of evidence helps to connect educational theory with practical diagnostic applications.

\par A key element is the progress achieved through immediate feedback on evaluations of these explanations during the analysis process. While research has examined tools in medical learning~\cite{gu2023augmenting,ouyang2023leveraging}, medical education still largely depends on the traditional ``\textit{mentor-apprentice}'' model to support students' professional growth~\cite{rassie2017apprenticeship,bleakley2011medical}. A future direction for \textit{e-MedLearn} is to integrate feedback into the analysis process, promoting the use of evidence as an explanation to enhance both student motivation and clinical practice.

\subsection{Building Sustainable Medical Learning Ecosystem}
\par The design of \textit{e-MedLearn} has proven effective in supporting individual learners in enhancing their PBL practice. Feedback also highlighted the need for shared discussion and participation with the system to facilitate collaborative learning. This aligns with the awareness that the essence of learning, both in content and process, lies in the acquisition of accumulated experience~\cite{kolb2014experiential}. These insights suggest opportunities to establish and sustain a medical education platform ecosystem that fosters continuous learner development and improvement. A key consideration is integrating past reflections into individual analyses and decision-making. Potential solutions include incorporating meta-information of cases, such as learner tags and case ratings, on the online learning platform to provide references and encourage active engagement.

\par Another key consideration is establishing a sustainable educational development process. Currently, medical cases are primarily compiled through casebook collections~\cite{singh2012100,baliga2012250}, which poses challenges when manually integrating these cases into the system workflow. Enhancing sustainability could be achieved by enabling instructors to upload cases directly to the platform, alongside automated crawling or data extraction from provided sources. This would streamline case integration and significantly contribute to the ongoing training and development of medical students.

\subsection{Incorporating Humanistic Factors Diagnosis for Comprehensive Diagnosis Consideration}
\par The design considerations for \textit{e-MedLearn} primarily emphasize practicing differential diagnosis within the PBL framework. While accurate diagnosis is crucial, medical practice also encompasses clinical decision-making, including treatment and recovery planning~\cite{o2010shifting,van2016decision}. Additionally, in real-world medical practice, especially for chronic diseases, patient management should not solely depend on life expectancy. It is essential to consider the patient's overall health, psychological state, and social context. Comprehensive medical care should therefore integrate both medical science and humanistic approaches. Training and expectations for physicians should align with the biopsychosocial model~\cite{engel1977need,borrell2004biopsychosocial}, which addresses biological, psychological, and social factors. Consequently, the PBL process should evolve to include not just clinical reasoning and differential diagnosis, but also value-based treatment planning and decision-making. In response to practice-driven needs, we aim to incorporate features into \textit{e-MedLearn} that encourage users to consider patient-centered factors during diagnosis and treatment. These enhancements will improve learners' ability to apply their knowledge in a holistic medical practice.

\subsection{Limitation and Future Work}
\par Our work has several limitations. \revision{First, In \textbf{Phase One}, our discussion sessions involved four physicians from two specialties. While the primary goal was to explore the current practices and challenges in PBL activities from the perspectives of both teachers and students, we acknowledge that the process of identifying issues was limited by the small sample size and the specialty-specific constraints. Future research with a broader scope and sustained efforts could offer more comprehensive insights into potential system improvements in the PBL process.} Second, the participants in this study were \revision{relatively few} and primarily from the orthopedic and neurosurgical fields. \revision{Although limiting the study to specific medical specialties reduces variability and allows for a focused evaluation of the system, thereby establishing a foundational understanding of how \textit{e-MedLearn} supports clinical reasoning in specialized medical contexts, the study does not account for potential differences between specialties that could influence the analysis and results. Future research should expand to include a broader range of clinical scenarios across a larger sample size and various medical specialties to provide a more comprehensive understanding of \textit{e-MedLearn}'s generalizability and effectiveness.} Third, as discussed in \autoref{testing_interviews}, learning and improvement are gradual processes, and the development of methodological skills requires sustained practice for effective assessment. We used the system's design concepts as a probe to promote targeted PBL practice among participants, offering further insights for design refinement. A longitudinal study is also planned to gather data on learners' long-term usage and feedback. Lastly, our current data is limited to hospitals affiliated with the collaborating physicians. In real-world scenarios, challenges such as cross-institutional or cross-platform data sharing may arise, along with ethical concerns about the use of unreviewed data. While addressing these issues is beyond the scope of our system's current design, we recognize their significance and are considering advocating for the establishment of institutions to supervise and review data-sharing practices, with the aim of fostering a sustainable health data ecosystem.

\section{Conclusion}
\par In this study, we explored the integration of clinical reasoning-based differential diagnosis to enhance targeted learning and improvement in PBL. The research followed a three-phase approach. In \textbf{Phase One}, we conducted interviews with physicians involved in PBL, gathering insights into their perceptions of the current PBL process, and areas for improvement. In \textbf{Phase Two}, we collaborated with experts to engage in design brainstorming. Through multiple design iterations, we identified key features that strengthen the application of clinical reasoning in PBL and developed the learner-centered \textit{e-MedLearn} system to support targeted continuous improvement. In \textbf{Phase Three}, we conducted a controlled study and testing interviews to collect the system's feedback on clinical diagnostic reasoning in PBL. The results indicated that \textit{e-MedLearn} effectively bridges the gap between knowledge acquisition and clinical practice, helping learners transition toward clinical diagnosis. 
Overall, \textit{e-MedLearn} provides valuable insights into the application of clinical diagnostic reasoning in PBL, with opportunities for further improvement, such as the development of a collaborative online platform and the integration of clinical scenario factors and medical humanities into patient care and treatment.

\begin{acks}
\par We thank anonymous reviewers for their valuable feedback. This work is supported by grants from the National Natural Science Foundation of China (No. 62372298), Shanghai Engineering Research Center of Intelligent Vision and Imaging, Shanghai Frontiers Science Center of Human-centered Artificial Intelligence (ShangHAI), and MoE Key Laboratory of Intelligent Perception and Human-Machine Collaboration (KLIP-HuMaCo).
\end{acks}

\bibliographystyle{ACM-Reference-Format}
\bibliography{sample-base}

\appendix

\onecolumn
\section{Criteria for the difficulty of cases}
\begin{table*}[h]
  \centering
  \caption{The difficulty criteria of cases from three dimensions: Incidence Rate, Logical Chain Length, and Relevance to Classic Diseases.}
  \begin{tabular}{p{3cm}p{3.5cm}p{4cm}p{4cm}}
\toprule
\textbf{Dimension} & \textbf{Low Difficulty} & \textbf{Medium Difficulty} & \textbf{High Difficulty} \\
\midrule
Incidence Rate & >100/100,000 annually & Medium 10-100/100,000 annually & <10/100,000 annually \\
\midrule
Logical Chain Length & 1-3 steps & 4-7 steps & >7 steps \\
\midrule
Relevance to Classic Diseases & Directly related & Somewhat related & Not related \\
\bottomrule
  \end{tabular}
  \label{tab:criteria}
\end{table*}

\section{Prompting Questions Template}
\begin{table*}[h]
  \centering
  \caption{Prompt questions template categorized by facts, ideas, learning issues, and action plan.}
  \label{tab:prompt_questions}
  \begin{tabular}{p{3cm}p{12cm}}
\hline
\textbf{Category} & \textbf{Questions} \\ \hline

\multirow{3}{*}{\textbf{Facts}} 
& Step 3: What facts are described in the current case? How can the <facts> be described in clinical syndrome descriptions? \\ \cline{2-2}
& Step 4: What results do the <updated data> present, and what facts and findings do they indicate?\\ \cline{2-2}
& Step 5: What key points about prognosis are mentioned in the case? Are the <facts> related to diagnosis and treatment consistent with your previous analysis? What inconsistencies or overlooked aspects exist?\\ \hline

\multirow{3}{*}{\textbf{Ideas}} 
& Step 3: Based on the fact that the patient exhibits <symptoms>, what inferences and hypotheses do you have? \\ \cline{2-2}
& Step 4: What can we infer from <findings>, and how do they help you update the problem representation and reasoning? \\ \cline{2-2}
& Step 5: What caused the inconsistencies and overlooked of <facts>? Which part of the constructed logical chain caused these problems? \\ \hline

\multirow{3}{*}{\textbf{Learning Issues}} 
& Step 3: / \\ \cline{2-2}
& Step 4: Are the <descriptions> of the patient consistent with your problem representation? Is the <updated data> consistent with your previous action plans?  \\ \cline{2-2}
& Step 5: What aspects were previously overlooked and needs consideration during <prognosis>? What was learned and strengthened from the analysis? What issues were identified and need further improvement? \\ \hline

\multirow{3}{*}{\textbf{Action Plan}} 
& Step 3: According to the <inferences and hypotheses>, what should the next action plans be? You may consider conducting specific examinations or treatments. \\ \cline{2-2}
& Step 4: Based on the above analysis and reasoning, what further action plans should be considered? You may consider conducting specific examinations or treatments. \\ \cline{2-2}
& Step 5: What are the targeted plans for further improvement and practice identified from the analysis?  \\ \hline
  \end{tabular}
\end{table*}

\section{Case Analysis Report Template}
\begin{table*}
\caption{\revision{Case analysis report template based on Clinical Reasoning Cycle (CRC).}}
\label{tab:report_template}
\begin{tabular}{p{9.25cm}p{7.75cm}}
\hline
\textbf{Step} & \textbf{Description} \\
\hline
\hline

\textbf{1. Consider the Patient Situation} & 
\textbf{Patient Background Information:}  
    - Age, gender, occupation, lifestyle, etc. \\
    - Chief Complaint: The main symptoms and concerns presented by the patient. \\
    - Medical History: Relevant diseases, treatments, or surgeries. \\
    
    \textbf{Clinical Context:}  
    - Clinical background and the patient’s current situation. \\
    - Potential urgency or severity of the health problem. \\
    
    \hline
    
    \textbf{2. Collect Cues and Information} & 
    \textbf{Clinical Data Collected:}  
    - Symptoms: Main complaints, nature of symptoms (pain location, intensity, etc.). \\
    - Signs: Physical exam findings (vital signs, physical exam). \\
    - Laboratory and Imaging Data: Blood tests, imaging (CT, MRI, etc.). \\
    - Other Information: Psychological status, lifestyle, medications, etc. \\
    
    \textbf{Information Summary:}  
    - List the key information gathered and summarize its relevance to the diagnosis. \\
    
    \hline
    
    \textbf{3. Process Information} & 
    \textbf{Analysis of Information:}  
    - Analyze the data collected and identify key clinical cues. \\
    - Interpret relationships among symptoms, signs, and test results. \\
    - Develop diagnostic hypotheses based on this analysis. \\
    
    \textbf{Reasoning Process:}  
    - Derive the most likely diagnoses from symptoms, signs, and test results. \\
    - Compare different diagnostic hypotheses and explain the rationale for each one. \\
    
    \hline
    
    \textbf{4. Identify Problems/Issues} & 
    \textbf{Core Issues:}  
    - Identify the main clinical problems or issues at hand. \\
    
    \textbf{Differential Diagnosis:}  
    - List possible diagnoses, both common and rare. \\
    - Provide evidence supporting each diagnosis and explain why others are excluded. \\
    
    \hline
    
    \textbf{5. Establish Goals} & 
    \textbf{Short-term Goals:}  
    - Determine immediate issues to address (e.g., pain relief, stabilize vital signs). \\
    - Set clear, actionable treatment goals (e.g., control blood pressure, further diagnostics). \\
    
    \textbf{Long-term Goals:}  
    - Set long-term health goals (e.g., stroke prevention, improve quality of life). \\
    
    \hline
    
    \textbf{6. Take Action} & 
    \textbf{Clinical Action Plan:}  
    - Based on the established goals, outline treatment plan including medications, further tests, and patient education. \\
    - Provide evidence supporting each action (e.g., clinical guidelines, literature). \\
    
    \textbf{Treatment Plan:}  
    - Outline the treatment steps, including medication choices or interventions. \\
    - Specify if further testing or referrals are required (e.g., neurology evaluation, imaging tests). \\
    
    \hline
    
    \textbf{7. Evaluate Outcomes} & 
    \textbf{Evaluate Treatment Effectiveness:}  
    - Monitor patient responses to treatment and assess improvements or side effects. \\
    
    \textbf{Assessment Criteria:}  
    - Evaluate based on symptom relief, lab results, and overall improvement. \\
    - Reflect on whether the treatment plan achieved the desired goals and whether adjustments are necessary. \\
    
    \hline
    
    \textbf{8. Reflect on the Process and New Learning} & 
    \textbf{Self-Reflection:}  
    - Reflect on potential biases or weaknesses in your reasoning process. \\
    - Summarize new knowledge gained from the case and how it can be applied in future scenarios. \\
    
    \textbf{Future Improvements:}  
    - Suggest improvements for the reasoning process or decision-making, to increase accuracy and efficiency. \\
    \hline
\end{tabular}
\end{table*}

\newpage
\section{Case Analysis Report Criteria}
\begin{table*}[h]
\centering
\caption{\revision{Case analysis report criteria based on Clinical Reasoning Cycle (CRC).}}
\label{tab:report_criteria}
\begin{tabular}{p{5cm}p{11cm}}
\hline
\textbf{Dimension} & \textbf{Grading Criteria: Description} \\
\hline
\textbf{1. Consider the Patient Situation} & \textbf{(1) Description of Patient Situation:} Does the student describe the patient's condition, including chief complaint, symptoms, and background in a clinical context? \\
\cline{2-2}
 & \textbf{(2) Extraction of Key Information:} Does the student identify key issues, highlighting symptoms or signs relevant to the diagnosis? \\
\hline
\textbf{2. Collect Cues and Information} & \textbf{(3) Completeness of Information Collection:} Has the student gathered relevant clinical information (history, exam, labs)? Is it comprehensive? \\
\cline{2-2}
 & \textbf{(4) Organization and Analysis of Information:} Is the information well-organized and integrated for analysis? \\
\hline
\textbf{3. Process Information} & \textbf{(5) Depth of Information Processing:} Does the student analyze the information and extract diagnostic clues? Is reasoning clear? \\
\cline{2-2}
 & \textbf{(6) Logical Consistency of Reasoning:} Does the student explain relationships between data points in a logical manner? \\
\hline
\textbf{4. Identify Problems/Issues} & \textbf{(7) Accuracy of Problem Identification:} Does the student accurately identify the core problem and explain differential diagnoses? \\
\cline{2-2}
 & \textbf{(8) Completeness of Differential Diagnosis:} Has the student proposed multiple diagnoses based on available data? \\
\hline
\textbf{5. Establish Goals} & \textbf{(9) Reasonableness of Goal Setting:} Does the student set clear goals based on identified issues and the patient's needs? \\
\cline{2-2}
 & \textbf{(10) Specificity and Feasibility of Goals:} Are the treatment goals specific, actionable, and achievable (e.g., symptom improvement, ruling out conditions)? \\
\hline
\textbf{6. Take Action} & \textbf{(11) Appropriateness of Actions Taken:} Has the student developed an appropriate plan (e.g., treatment, further tests, follow-up)? \\
\cline{2-2}
 & \textbf{(12) Evidence Support for Actions:} Does the student provide strong supporting evidence for chosen actions (e.g., literature, guidelines)? \\
\hline
\textbf{7. Evaluate Outcomes} & \textbf{(13) Comprehensiveness of Outcome Evaluation:} Does the student evaluate the outcomes comprehensively, considering treatment effects and patient responses? \\
\cline{2-2}
 & \textbf{(14) Reflection and Plan Adjustment:} Does the student reflect on the outcomes and adjust the plan if necessary? \\
\hline
\textbf{8. Reflect on the Process and New Learning} & \textbf{(15) Depth of Reflection and Self-assessment:} Does the student self-reflect on strengths and weaknesses in their reasoning process? \\
\cline{2-2}
 & \textbf{(16) Integration and Application of Learning:} Can the student apply new learning to improve future clinical reasoning? \\
\hline
\end{tabular}
\end{table*}

\begin{itemize}
    \item \textbf{7 points:} Completely meets the standard; information is collected thoroughly, analysis is logically clear, and reasoning is well-supported.
    \item \textbf{5-6 points:} Mostly meets the standard, but certain aspects could be more detailed or further developed.
    \item \textbf{3-4 points:} There are some omissions or insufficient areas; reasoning or analysis lacks comprehensiveness or rigor.
    \item \textbf{1-2 points:} Significant omissions or errors; reasoning is incoherent or lacks sufficient evidence.
    \item \textbf{0 points:} Does not meet the requirements at all; critical components are missing or clinical reasoning standards are not followed.
\end{itemize}

\section{In-task Questionnaire in Testing Interviews}
\begin{table}[H]
  \centering
  \caption{In-task questionnaire assessing feature usefulness.}
  \label{tab:in_task_questionnaire}
  \begin{tabular}{lp{0.6\textwidth}}
    \toprule
    \textbf{Category} & \textbf{Question} \\
    \midrule
    \multirow{4}{*}{Feature Usefulness} 
    & 1. The diagnostic list is useful for supporting evidence-based analysis. \\
    & 2. The QA section helps me facilitate the exploration of evidence. \\
    & 3. The prompting questions assist in structuring my clinical reasoning. \\
    & 4. The mind map is useful for organizing my thoughts during analysis. \\
    \bottomrule
  \end{tabular}
\end{table}


\clearpage
\section{Prompt Template}
\begin{table}[ht]
\centering
\caption{\revision{Prompt template for the QA section in the \textit{e-MedLearn} interface.}}
\renewcommand{\arraystretch}{1.5}
\begin{tabular}{p{0.9\textwidth}}
\hline
\textbf{Prompt} \\
\hline
You are a medical assistant. Based on the following patient case data, respond to the user's symptom inquiries by adhering to the following guidelines: \\

1. \textbf{Single Symptom Evaluation:} \\
    - If the patient's case data includes the mentioned symptom, respond with "Yes" and provide relevant details (e.g., severity, duration, related measurements). \\
    - If the symptom is not present in the case data, respond with "Irrelevant". \\

2. \textbf{Multiple Symptoms Evaluation:} \\
    - If the user mentions multiple symptoms, evaluate each symptom individually. \\
    - Present the results in a clear, organized list format. For example: \\
        Symptom 1: Yes \\
        Relevant Data: [Details] \\
        
        Symptom 2: Irrelevant \\
        
        Symptom 3: Yes \\
        Relevant Data: [Details] \\

3. \textbf{Response Format:} \\
    - Ensure responses are clear, concise, and professionally formatted. \\
    - Use bullet points or numbered lists for multiple symptoms to enhance readability. \\

4. \textbf{Language and Tone:} \\
    - Maintain a professional and objective tone. \\
    - Avoid subjective judgments or ambiguous language. \\

5. \textbf{Error Handling:} \\
    - If the user's input is unclear or does not specify a symptom, respond with a clarifying question. For example: "Could you please specify the symptom you are inquiring about?" \\

\textbf{Example Scenario:} \\

- User Input: "Does the patient have a fever and headache?" \\

- Possible Response: \\
    Fever: Yes \\
    Relevant Data: Temperature is 38.5°C, duration of 3 days. \\
    
    Headache: Irrelevant \\

Ensure that all responses strictly follow the above guidelines to maintain consistency and reliability. \\

\hline
\end{tabular}
\label{tab:structured_prompt}
\end{table}

\clearpage
\section{Initial Prototype}
\begin{figure*}[h]
\centering
\includegraphics[width=0.7\textwidth]{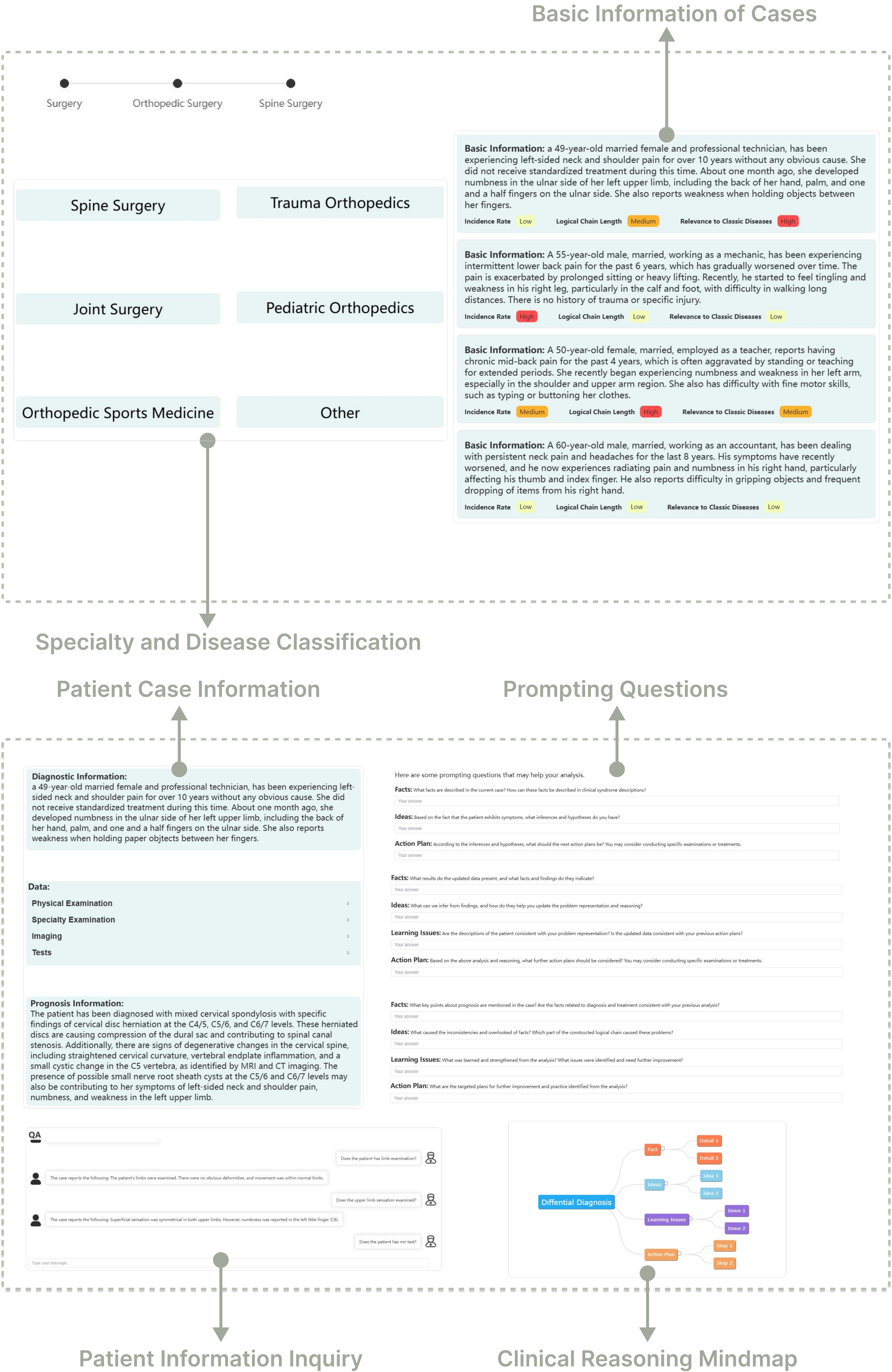}
\caption{\revision{The initial version of the \textit{e-MedLearn} prototype.}}
\label{fig:prototype}
\end{figure*}

\clearpage
\section{Baseline System}
\begin{figure*}[h]
\centering
\includegraphics[width=\textwidth]{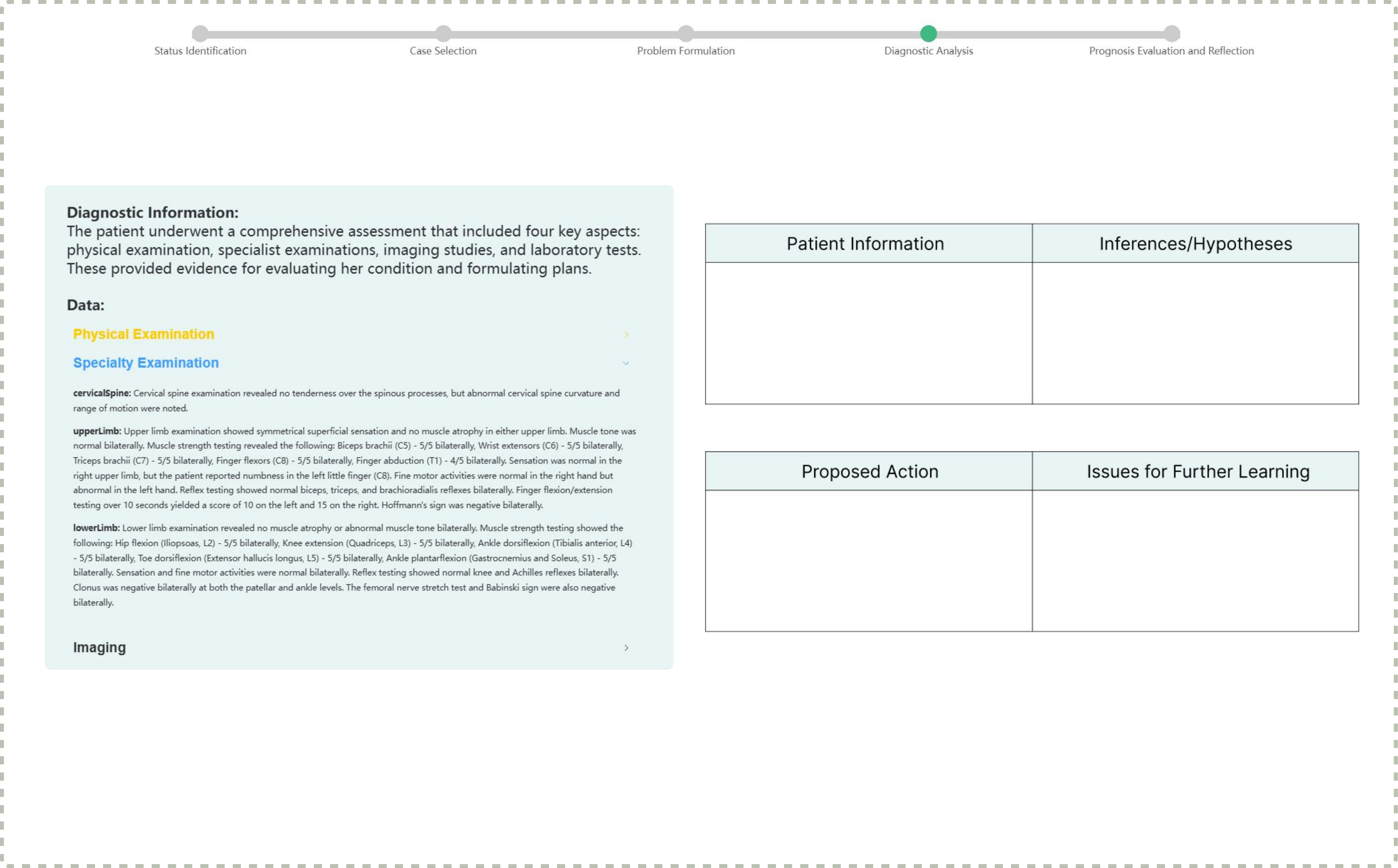}
\caption{\revision{The interface of baseline system in the controlled study.}}
\label{fig:baseline}
\end{figure*}

\clearpage
\section{System Interface}
\begin{figure*}[h]
\centering
\includegraphics[width=0.9\textwidth]{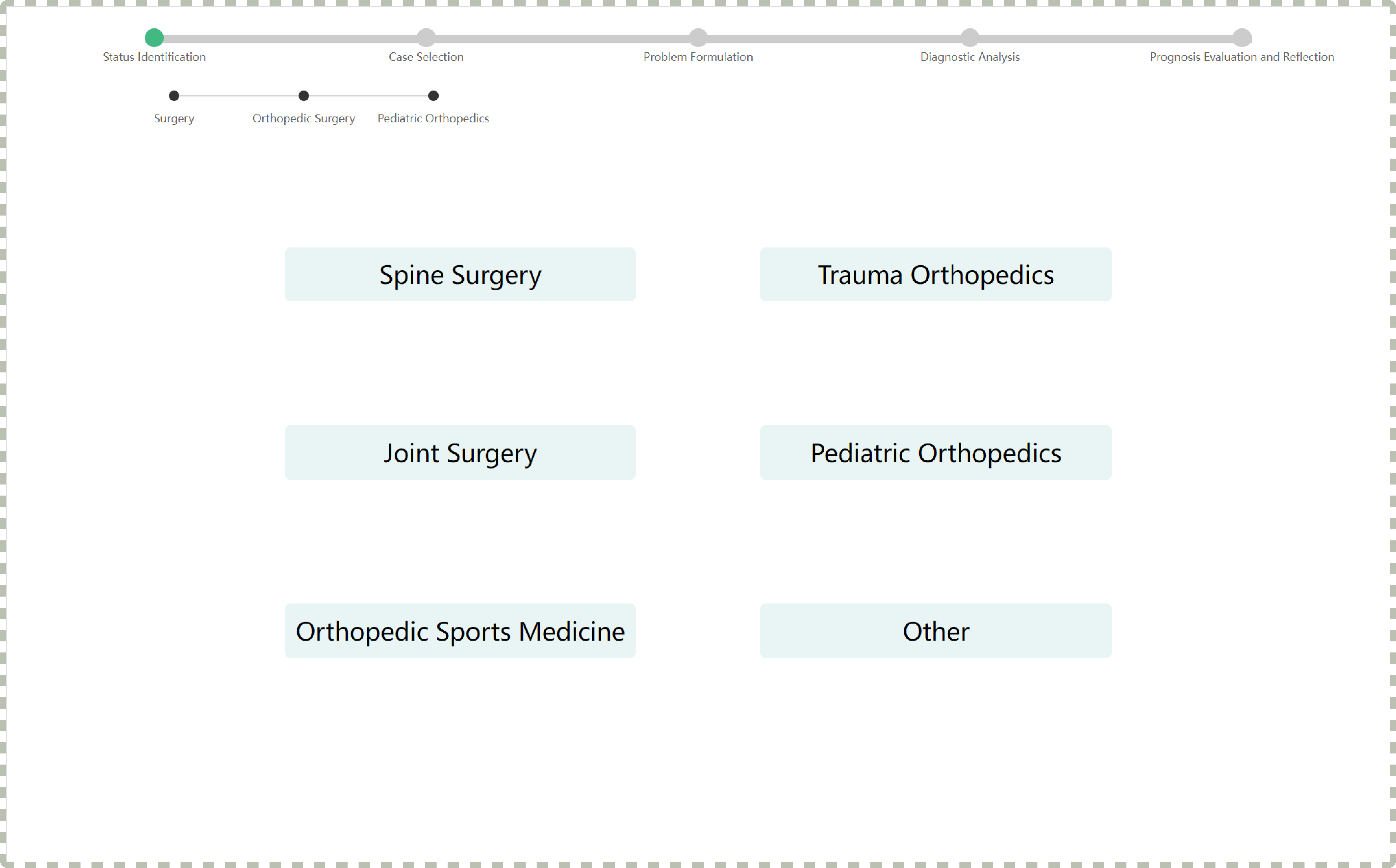}
\caption{Step1: Status Identification}
\label{fig:appendix_1}
\end{figure*}

\begin{figure*}[h]
\centering
\includegraphics[width=0.9\textwidth]{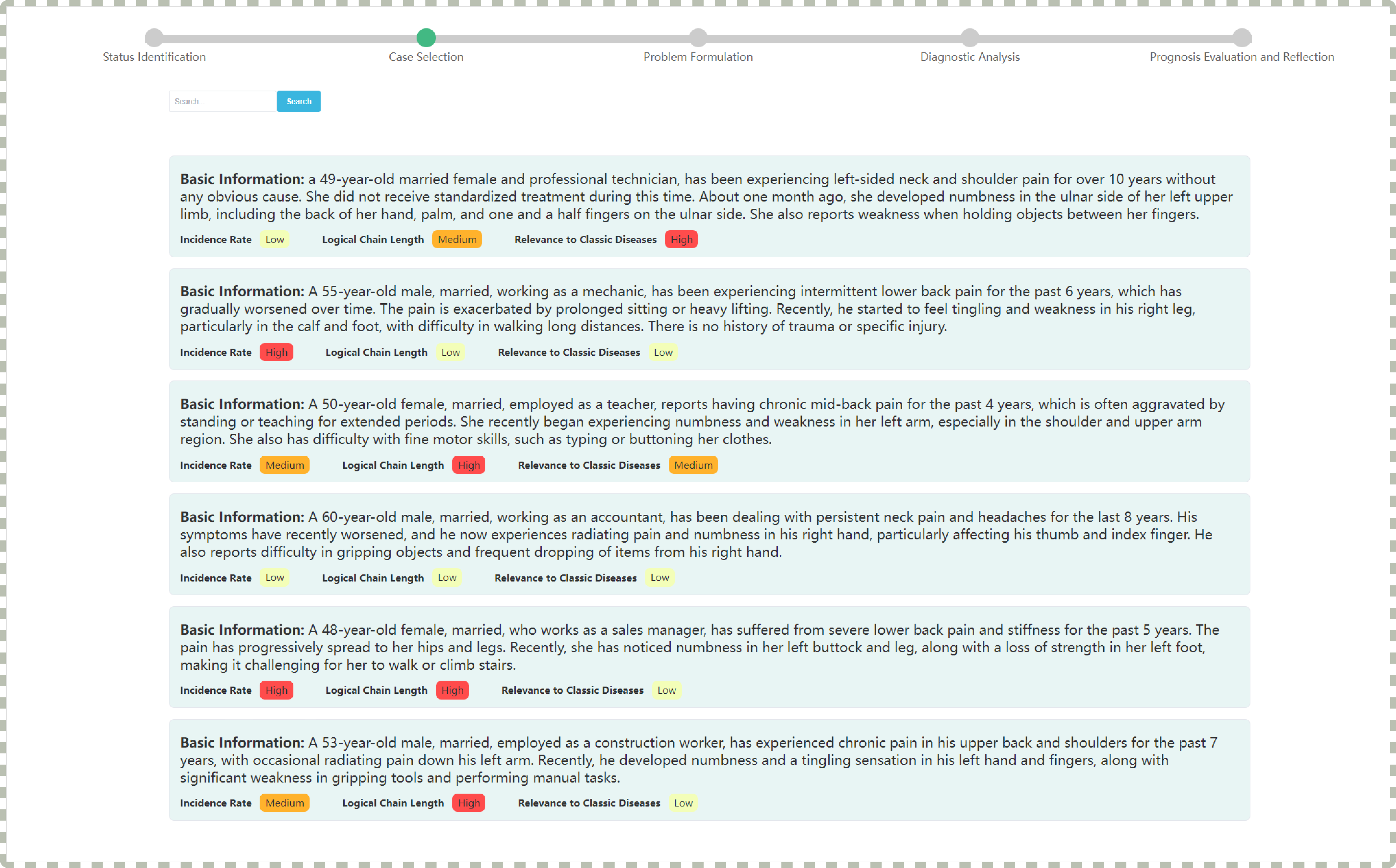}
\caption{Step2: Case Selection}
\label{fig:appendix_2}
\end{figure*}

\begin{figure*}[h]
\centering
\includegraphics[width=0.9\textwidth]{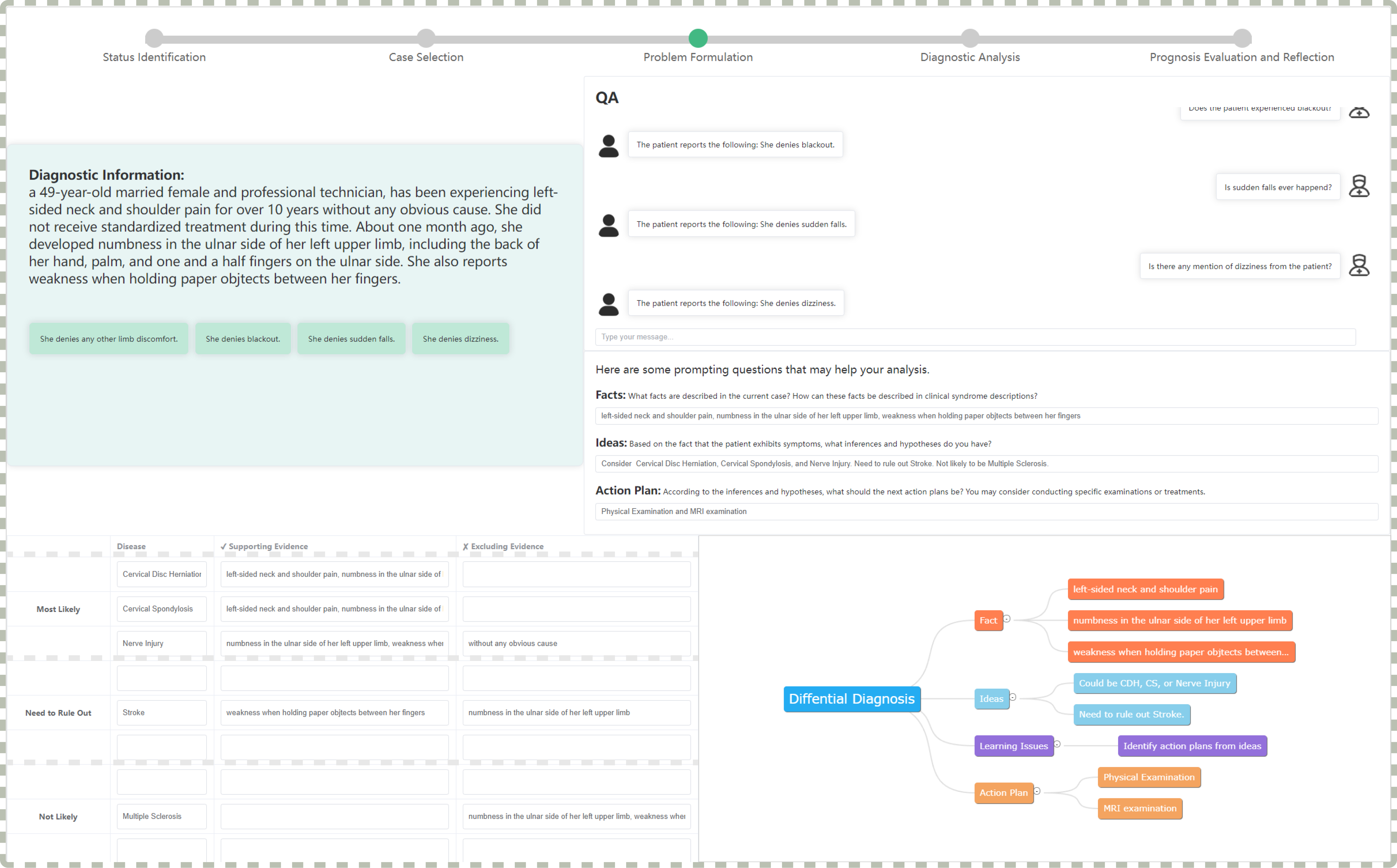}
\caption{Step3: Problem Formulation}
\label{fig:appendix_3}
\end{figure*}

\begin{figure*}[h]
\centering
\includegraphics[width=0.9\textwidth]{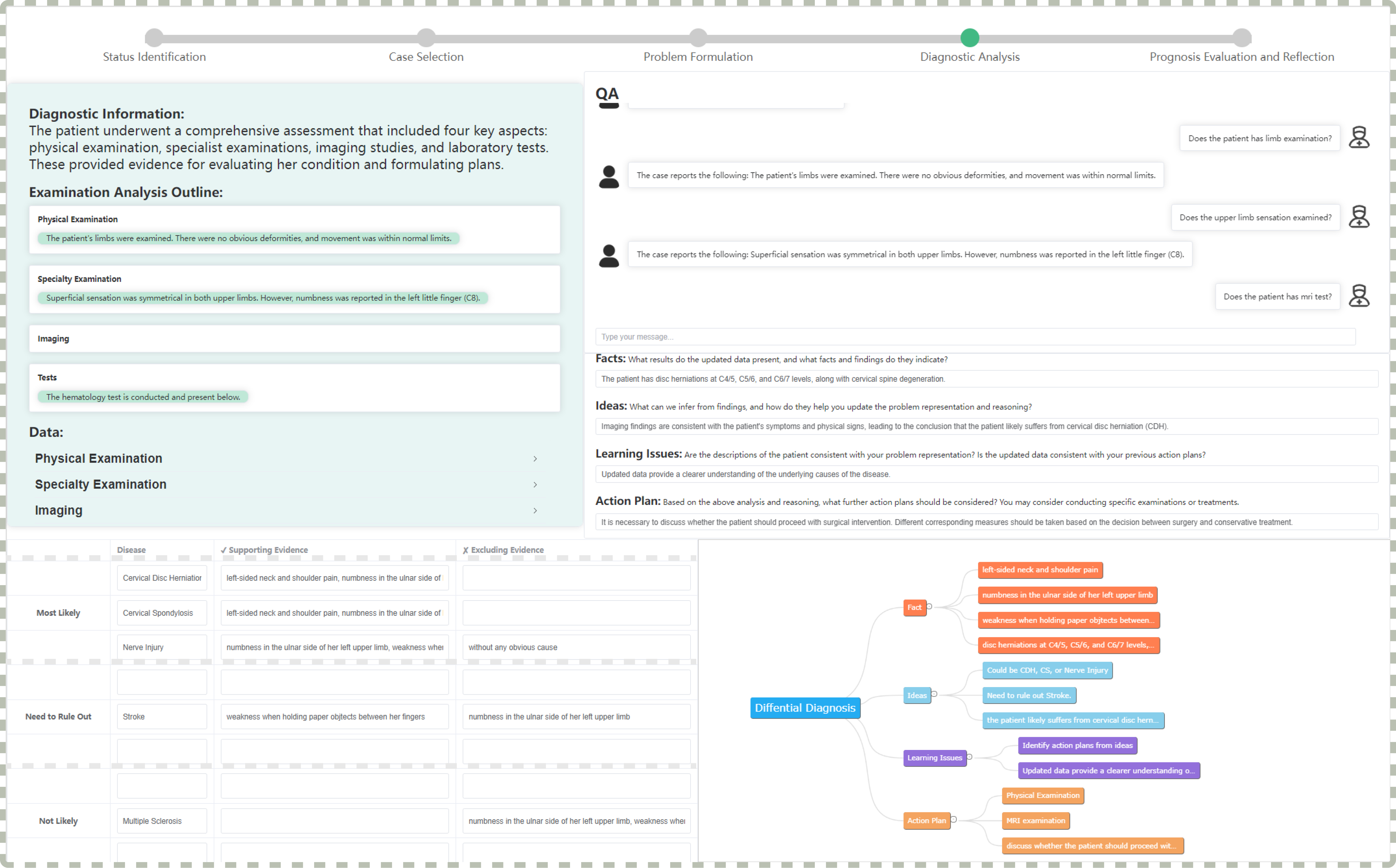}
\caption{Step4: Diagnostic Analysis}
\label{fig:appendix_4}
\end{figure*}

\begin{figure*}[h]
\centering
\includegraphics[width=0.9\textwidth]{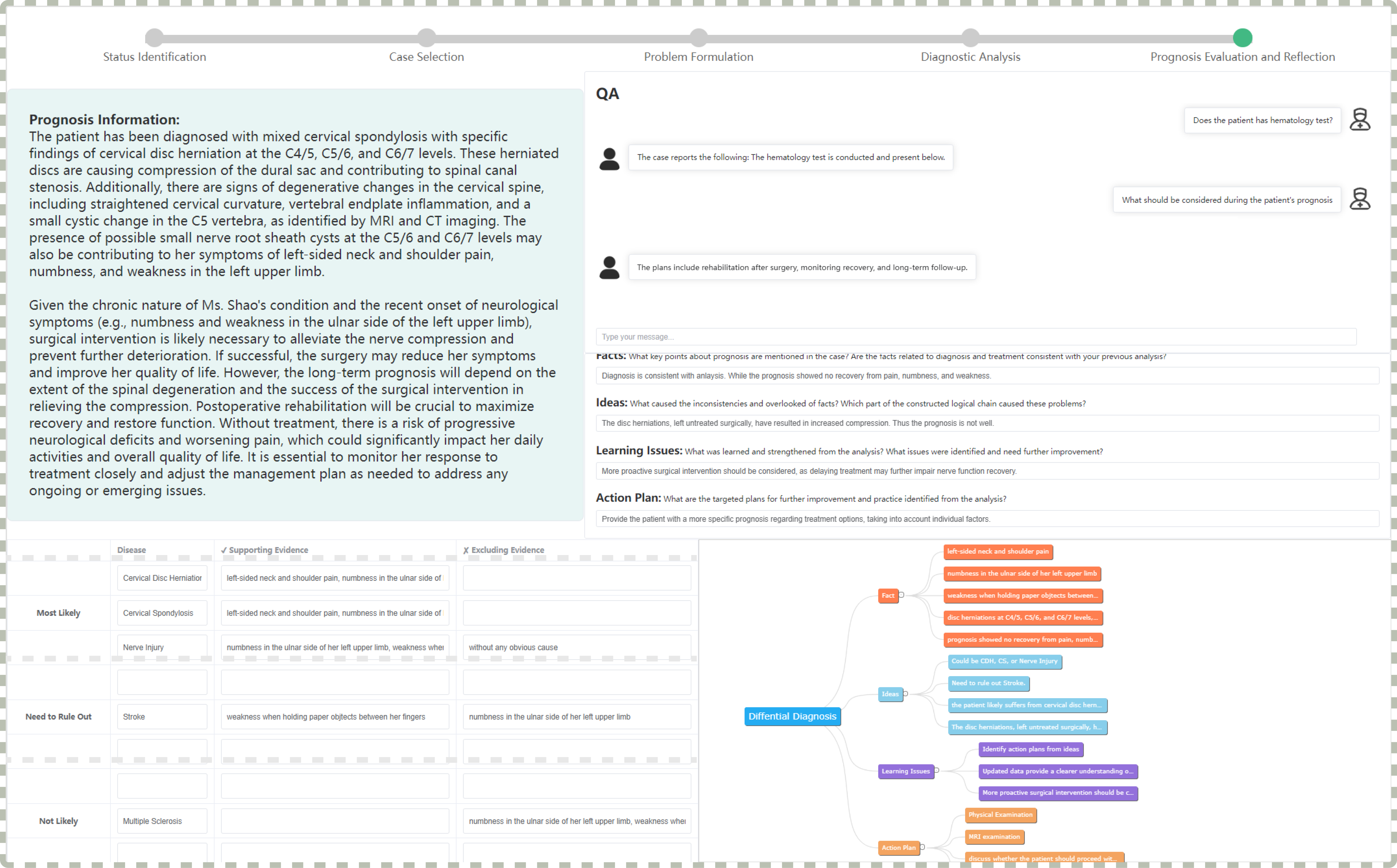}
\caption{Step5: Prognosis Evaluation and Reflection}
\label{fig:appendix_5}
\end{figure*}

\clearpage

\section{System Walkthrough}
\begin{figure*}[h]
\centering
\includegraphics[width=\textwidth]{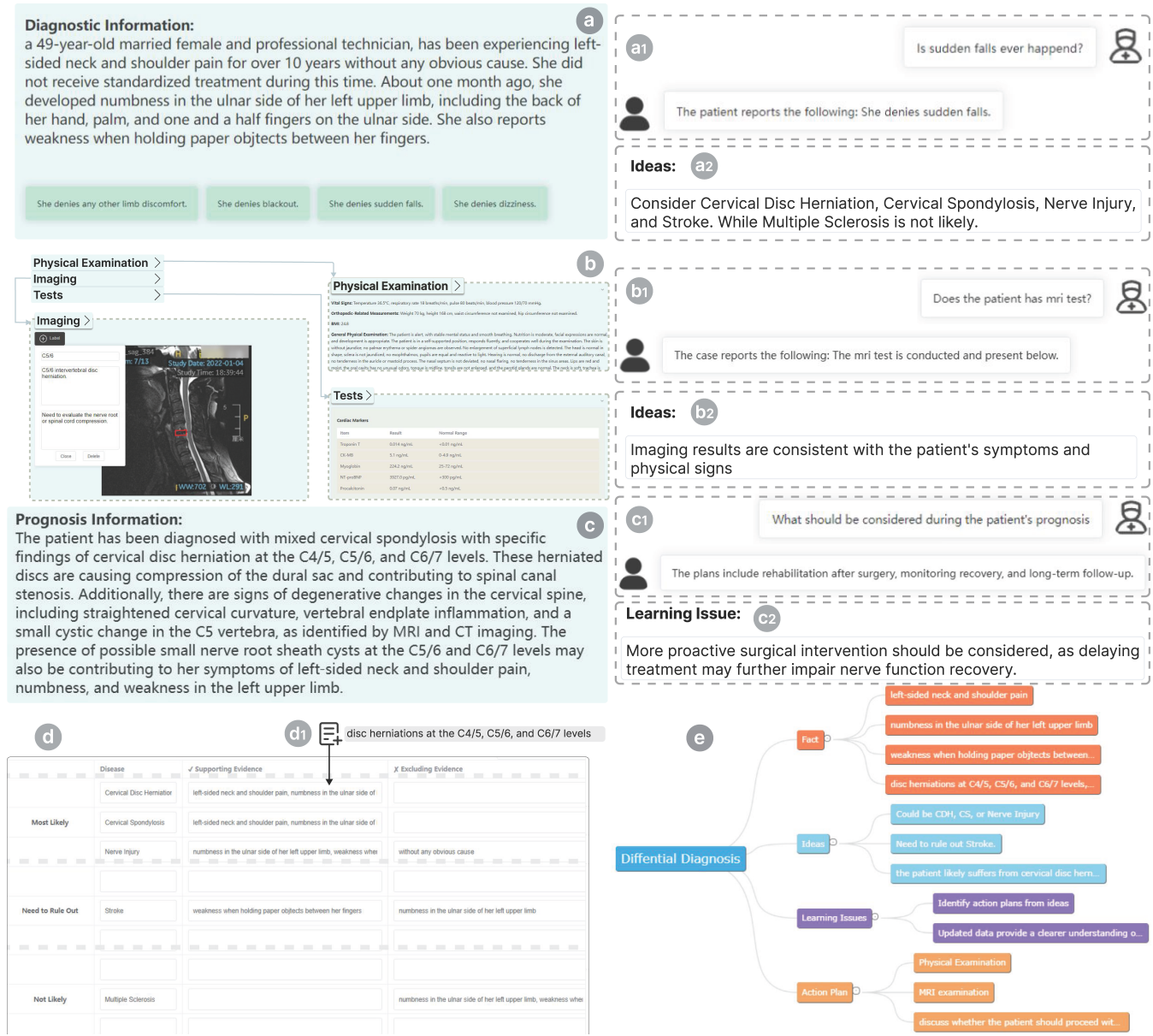}

\caption{\revision{The process of conducting case analysis with \textit{e-MedLearn} after \textbf{Stage 1: Case Selection} consists of (a) Presenting the preliminary case data, exploring the evidence, and formulating initial ideas of diagnosis. (b) Perform diagnostic analysis using data from examinations or tests and update the ideas of the diagnosis. (c) Consider prognosis and assess analysis results, reflect ideas to conclude learning issues. (d) Record and update the diagnosis in the diagnosis list. (e) Record and review the insights and thoughts of the analysis process in the mind map.}}
\label{fig:walkthrough}
\end{figure*}

\par \revision{We present the system's interactive usage through a complete walkthrough. In this scenario, Jon, a medical intern, uses \textit{e-MedLearn} for his learning process. Initially, he selected the orthopedic category with a focus on spine-related diseases as his area of interest. Jon then selected a case with a low \textit{incidence rate} but significant \textit{relevance to classic diseases} for analysis.}

\par \revision{In \textit{Step 3: Problem Formulation}, Jon first reviewed the initial diagnostic information of the case (\autoref{fig:walkthrough}-a). He then engaged with the system in the QA section to further explore and collect evidence (\autoref{fig:walkthrough}-a1). Through questioning, Jon confirmed that the patient \textit{denies experiencing blackouts, sudden falls, dizziness, or any other limb discomfort}.}

\par \revision{Following prompting questions, Jon organized the collected information: \textit{the current facts indicate that the patient experiences left-sided neck and shoulder pain, numbness in the ulnar side of her left upper limb, and weakness when holding paper objects between her fingers}. Based on these facts, the ideas of possible hypotheses and inferences include: \textit{Consider Cervical Disc Herniation, Cervical Spondylosis, Nerve Injury, and Stroke. While Multiple Sclerosis is not likely} (\autoref{fig:walkthrough}-a2). The proposed action plans involve conducting a \textit{Physical Examination and an MRI examination}.}

\par \revision{Jon then updated the preliminary diagnoses in the diagnosis list (\autoref{fig:walkthrough}-d) accordingly: \textit{Based on supporting evidence, the most likely diagnoses are Cervical Disc Herniation and Cervical Spondylosis. In addition, Stroke needs to be ruled out, as weakness when holding paper objects between her fingers serves as supporting evidence for this possibility. Multiple Sclerosis is considered unlikely because the numbness in the ulnar side of her left upper limb and weakness when holding paper objects provide evidence against this diagnosis}.}

\par \revision{In \textit{Step 4: Diagnostic Analysis}, Jon further inquired in the QA section to identify the types of updated examination and tests (\autoref{fig:walkthrough}-b1), and then reviewed the detailed results (\autoref{fig:walkthrough}-b). Following the prompting questions, Jon analyzed the data and extracted the following facts: \textit{The patient has disc herniations at the C4/5, C5/6, and C6/7 levels, along with cervical spine degeneration.}}

\par \revision{These findings led him to the idea that \textit{imaging results are consistent with the patient's symptoms and physical signs} (\autoref{fig:walkthrough}-b2), leading to the conclusion that the patient likely suffers from \textit{Cervical Disc Herniation}. Jon noted that \textit{the updated data provides a clearer understanding of the underlying causes of the disease}. Following this, Jon updated the diagnosis list, adding the evidence provided by MRI examination for \textit{Cervical Disc Herniation} (\autoref{fig:walkthrough}-d1).}


\par \revision{In \textit{Step 5: Prognosis and Evaluation}, Jon reviewed the patient's prognosis information (\autoref{fig:walkthrough}-c) and gathered key considerations regarding the prognosis through inquiry in the QA section (\autoref{fig:walkthrough}-c1). Jon confirmed that the diagnosis results align with his previous analysis. Reflecting further, Jon realized that \textit{more proactive surgical intervention should be considered, as delaying treatment may further impair nerve function recovery} (\autoref{fig:walkthrough}-c2). The next action plan involves \textit{providing the patient with a more specific prognosis regarding treatment options and taking into account individual factors}.}

\end{document}